\documentclass[structabstract]{aa}  
\usepackage{graphicx}
\usepackage[]{natbib}
\usepackage{txfonts}
\newcommand{\teff}{T$_{\rm eff}$}
\newcommand{\logg}{$\log~g$}
\newcommand{\feh}{$\rm [Fe/H]$}
\newcommand{\met}{${\rm [M/H]}$}
\newcommand{\aabun}{${\rm [\alpha/Fe]}$}
\newcommand{\kms}{km~s$^{-1}$}
\newcommand{\vrad}{${\rm V_{rad}}$}

\begin{document}
   \title{The Gaia-ESO Survey: the Galactic Thick to Thin Disc transition\thanks{Based on observations collected with the FLAMES spectrograph at the VLT/UT2 telescope (Paranal Observatory, ESO, Chile), for the Gaia-ESO Large Public Survey, programme 188.B-3002}}

   \author{A. Recio-Blanco\inst{1}
          \and
          P. de Laverny\inst{1}
          \and
          G. Kordopatis\inst{2}
	  \and
          A. Helmi\inst{3}
          \and
	  V. Hill\inst{1}
          \and
          G. Gilmore\inst{2}
          \and
          R. Wyse\inst{4}
          \and
          V. Adibekyan\inst{5}
\and
S. Randich\inst{6}
\and
M. Asplund\inst{7}
\and
S. Feltzing\inst{8}
\and
R. Jeffries\inst{9}
\and
G. Micela\inst{10}
\and
A. Vallenari\inst{11}
\and
E. Alfaro\inst{12}
\and
C. Allende Prieto\inst{13}
\and
T. Bensby\inst{8}
\and
A. Bragaglia\inst{14}
\and
E. Flaccomio\inst{10}
\and
S. E. Koposov\inst{2,20}
\and
A. Korn\inst{15}
\and
A. Lanzafame\inst{16}
\and
E. Pancino\inst{14,17}
\and
R. Smiljanic\inst{18,19}
\and
R. Jackson\inst{9}
\and
J. Lewis\inst{2}
\and
L. Magrini\inst{6}
\and
L. Morbidelli\inst{6}
\and
L. Prinsinzano\inst{10}
\and
G. Sacco\inst{6}
\and
C. C. Worley\inst{2}
\and
A. Hourihane\inst{2}
\and
M. Bergemann\inst{2}
\and
M. T. Costado\inst{12}
\and
U. Heiter\inst{15}
\and
P. Joffre\inst{2}
\and
C. Lardo\inst{14}
\and
K. Lind\inst{2}
\and
E. Maiorca\inst{6}
          }

   \institute{
	 Laboratoire Lagrange (UMR7293), Universit\'e de Nice Sophia Antipolis, CNRS, Observatoire de la 
  	 C\^ote d'Azur, CS 34229,F-06304 Nice cedex 4, France\\   \email{arecio@oca.eu}
         \and
         Institute of Astronomy, Cambridge University, Madingley Road, Cambridge CB3 0HA, United Kingdom 
         \and
         Kapteyn Astronomical Institute, University of Groningen, PO Box 800, 9700 AV Groningen, The Netherlands
         \and
         Johns Hopkins University, Homewood Campus, 3400 N Charles Street, Baltimore, MD 21218, USA 
         \and
         Centro de Astrof\' isica da Universidade do Porto, Rua das Estrelas, 4150-762 Porto, Portugal 
         \and
         INAF-Osservatorio Astrofisico di Arcetri, Largo E. Fermi, 5, 50125, Firenze, Italy
         \and
         Research School of Astronomy and Astrophysics, Australian National University, Canberra, ACT 2611, Australia 
         \and
         Dept. of Astronomy and Theoretical physics, Lund university, Box 43, SE-22100 Lund, Sweden
         \and
         Astrophysics Group, Keele University, Keele, Staffordshire ST5 5BG, UK 
         \and
         INAF - Osservatorio Astronomico di Palermo, Piazza del Parlamento 1, 90134 Palermo, Italy
         \and
         Osservatorio Astronomico di Padova, Vicolo dell’Osservatorio 5, 35122 Padova, Italy 
         \and
         Instituto de Astrof\' isica de Andaluc\' ia (IAA-CSIC), Glorieta de la Astronom\' ia, E-18008-Granada, Spain 
         \and
         Instituto de Astrof\' isica de Canarias, E-38205 La Laguna, Tenerife, Spain 
         \and 
         INAF-Osservatorio Astronomico di Bologna, via Ranzani 1, Bologna, Italy 
         \and
         Department of Physics and Astronomy, Uppsala University, Box 516, SE-75120 Uppsala, Sweden 
         \and
         Dipartimento di Fisica e Astronomia, Sezione Astrofisica, Universit\'{a} di Catania, via S. Sofia 78, 95123, Catania, Italy
         \and
         ASI Science Data Center, Via del Politecnico SNC, I-00133 Roma, Italia
         \and
         European Southern Observatory, Karl-Schwarzschild-Str. 2, 85748 Garching bei M\"unchen, Germany
         \and
         Department for Astrophysics, Nicolaus Copernicus Astronomical Center, ul. Rabia\'nska 8, 87-100 Toru\'n, Poland
         \and
         Moscow M.V. Lomonosov State University, Sternberg Astronomical Institute, Universitetskij pr., 13, 119992 Moscow, Russia
             }

   \date{Received; accepted}

   \abstract
    {}   
    {The nature of the Thick Disc and its relation with the Thin Disc is presently an important subject of 
debate. In fact, the structural and chemo-dynamical transition between disc populations can be used as a test of the
proposed models of Galactic disc formation and evolution. }
    {We have used the atmospheric parameters, \aabun \ abundances  and radial velocities, determined from 
the Gaia-ESO Survey GIRAFFE spectra of FGK-type stars (first nine months of observations), to provide
a chemo-kinematical characterisation of the disc stellar populations. We focuss on a
subsample of 1016 stars with high quality parameters, covering the volume $|Z|<$4.5~kpc and R in the range 2-13~kpc.
}
    {We have identified a thin to thick disc separation in the \aabun \ v.s. \met \ plane, thanks to
the presence of a low-density region in the number density distribution. The thick disc stars seem to lie in progressively 
thinner layers above the Galactic plane, as metallicity increases and \aabun \ decreases. On the contrary, 
the thin disc population presents a constant value of the mean distance to the Galactic plane at all metallicities.
In addition, our data confirm the already known correlations between V$_{\phi}$ and \met \ for the two discs. 
For the thick disc sequence, 
a study of the possible contamination by thin disc stars suggests a gradient up to 64$\pm$9~km~s$^{-1}$~dex$^{-1}$.
The distributions of azimuthal velocity, vertical velocity, and orbital parameters are also analysed for the chemically
separated samples.
Concerning the gradients with galactocentric radius,  we find, for the thin disc, a flat behaviour of the azimuthal velocity, 
a metallicity gradient equal to -0.058$\pm$0.008~dex~kpc$^{-1}$ and a very small positive \aabun \ gradient.
For the thick disc, flat gradients in \met \ and \aabun \ are derived.
}
    {Our chemo-kinematical analysis suggests a picture in which the thick disc 
seems to have experienced a settling process, during which its rotation 
increased progressively, and, possibly, the azimuthal velocity dispersion decreased. At \met$\approx$-0.25~dex and 
\aabun$\approx$0.1~dex, the
mean characteristics of the thick disc in vertical distance to the Galactic plane, rotation, 
rotational dispersion and stellar orbits eccentricity are in agreement with that of the thin disc stars of the same metallicity,
suggesting a possible connection between these two populations at a certain epoch of the disc evolution.
Finally, the results presented here, based only on the first months of the Gaia ESO Survey 
observations, confirm how crucial are today large high-resolution spectroscopic surveys outside the solar neighbourhood for our understanding
of the Milky Way history.
} 


   \maketitle
%

\section{Introduction}
Understanding the chemodynamical evolution of the Milky Way disc 
stellar populations is a crucial step to reconstruct the history of
our Galaxy. This approach can
shed new light on the detailed processes that led the Milky Way to
form and evolve as a disc-type galaxy, in the more general context
of galaxy evolution.
In particular, since the first works highlighting the existence of the Milky Way's thin/thick disc 
dichotomy \citep{Yoshii,GilmoreReid}, the study of the structural and chemodynamical 
transition between the two disc populations has opened new pathways to constraint Galactic
evolution models. 
External mechanisms, supported by the hierarchical formation of galaxies
in the $\Lambda$CDM paradigm, have been invoked to explain the chemodynamical characteristics of thick disc stars.
Among them, the accretion of dwarf galaxies \citep[e.g.][]{Statler, Abadi} and minor mergers of satellites 
\citep[e.g.][]{Quinn, Villalobos} heating dynamically the disc, have been proposed. Similarly, \citet{Jones83}
and \citet{Brook1,Brook2} have proposed the accretion of a gas-rich merger from the collapse of which the thick disc
would have formed. On the other hand, purely internal formation mechanisms have also been proposed, like the early
turbulent phase of the primordial disc \citep[e.g.][]{Bournaud} and the stars radial migration due to resonances with the
spiral structure or the bar of the Milky Way \citep[e.g.][]{Schonrich2009,Minchev10,Loebman}.
The relative importance of the different proposed physical processes of evolution can now be tested
with increasing robustness, thanks to the Galactic spectroscopic surveys targeting the disc stars.

Spectroscopic studies of the disc populations were initially based on kinematically 
selected samples of solar neighbourhood stars \citep[e.g.][]{Prochaska2000,Fuhrmann04,Bensby05,Reddy06}, with some exceptions like 
\citet{Edvardsson93} or \citet{Fuhrmann98}. In the classical picture that emerged from those studies,
the thick disc population is characterized, compared to the thin disc one, as 
kinematically hotter \citep{CasettiDinescu2011}, and therefore, composed by stars
moving in Galactic orbits with a higher scale height \citep[$\sim$900~pc vs.  $\sim$300~pc,
e.g.][]{Juric2008}. In addition, the thick disc metallicity distribution peaks at lower values 
\citep[$\sim$-0.5~dex compared to $\sim$-0.2~dex for the thin disc, e.g][]{WyseGilmore95,Kordopatis11b,Lee2011b}.
With respect to the thin disc, the thick disc stars are older and have an enhanced ratio of $\alpha$-element
abundances over iron \citep[\aabun, e.g.][]{Bensby05,Bensby07,Fuhrmann08}. 
Nevertheless, it progressively appeared 
that the distributions in the above mentioned physical parameters displayed significant overlaps between
the thin and the thick disc populations \citep[e.g.][]{BensbyFeltzing10}. This blurred 
our comprehension of the interplay between the discs and questionned the classical
kinematically-based definitions \citep[][]{Bovy12Mono}.

On the other hand, as discussed
in \citet{Bovy2012a}, defining stellar populations by abundances patterns is a better
approach than the traditional kinematical criteria, as chemical abundances can correlate with disc
structure, but are formally independent of it. 
Moreover, the identification of possible chemical evolutionary paths and/or gaps in the abundance 
ratios distributions can be crucial to disentangle 
the otherwise overlapping populations in the kinematical space.

In the recent years, the number of stars analysed with high enough spectroscopic resolution
to provide detailed chemical diagnostics has increased from a few hundreds to several tens
of thousands. 
The advent of Milky Way spectroscopic
surveys and of automatised chemical analysis techniques have improved both the statistical robustness
and the homogeneity of the data as a consequence. From the theoretical side, new approaches based, at least partially,
on the chemical identification of disc sub-populations, as the chemical-tagging \citep{FreemanBH} or 
the mono-abundance populations \citep{RixBovy} methods, are opening promising pathways for constraining
the Milky Way's evolutionary processes. This is also challenging our previous conception of the Milky Way disc and
the way in which stellar populations can be defined.\\

In this context, the low resolution (R $\sim$ 2\,000) SEGUE spectroscopic survey addressed the question of the thin
to thick disc transition with a robust statistical approach, inside and outside the solar neighbourhood. 
\cite{Lee2011b} analysed $\sim$ 17\,300 G-type dwarfs with errors of about 0.23 dex in [Fe/H] \citep{Smolinski2011} and
0.1 dex for \aabun \ \citep{Lee2011a}. This was complemented by \cite{Schlesinger} who analysed 24\,270 G and 16\,847 K
dwarfs at distances from 0.2 to 2.3~kpc from the Galactic plane.
The \citet{Lee2011b} \aabun ~versus [Fe/H] distribution of number densities (see their Fig. 2) already 
motivated the authors to chemically separate the thin and the thick disc populations, although no clear gap is observed
between the two. The SEGUE data interpretations differ from authors proposing no
thin-thick disc distinction \citep{Bovy2012a} to those allowing the existence of a distinct thick-disc 
component formed through an external mechanism \citep{Liu}. Nevertheless, selection effects have been invoked to
explain some of the distinctions between the \citet{Lee2011a} and the \citet{Bovy2012a} chemical distributions.
On the other hand, the data interpretation in the sense of no thin-thick disc distinction agrees with the results 
of the \citet{Shonrich09NoThick} theoretical study.

More recently, \citet{Boeche} analysed the chemo-kinematical characteristics of 9\,131 giants included in the 
last RAVE survey data release, with available \met \ and \aabun. The RAVE survey data have a higher resolution
(R$=$7\,500) than SEGUE spectra, but with a wavelength range limited to the infrared Ca~II triplet. The \citet{Boeche}
analysis, based on the eccentricity-Zmax plane combined with additional orbital parameters,
allowed them to identify three stellar populations that could be associated with the Galactic thin disc, 
a dissipative component composed mostly of thick-disc stars, and the halo. Nevertheless, their thin and thick
disc populations, defined with the above mentioned dynamical criteria, do not clearly separate in their 
\aabun \ v.s. \met  \ plane (c.f. their Fig.~10).

On the other hand, recent studies of solar neighbourhood samples, analysed with high resolution spectroscopic
data, have revealed the existence of a gap in the \aabun ~versus [Fe/H] plane \citep[e.g][]{Fuhrmann04,Reddy06,Bensby07}. Recently,
\citet{Adibekyan2013} used the stellar sample of 1\,111 long-lived FGK dwarf stars from \citet{Adibekyan2012} 
to separate and characterise the different Galactic stellar subsystems, taking into account the existence of
the above mentioned gap. Their \aabun ~measurements are based on 
Mg, Si and Ti averaged abundances and the typical relative uncertainties in the metallicity and \aabun ~are of 
about 0.03 dex. \citet{Haywood2013} derived ages of the \citet{Adibekyan2012} sample,
concluding that two regimes appear in the age-\aabun ~plane, that they identify as the epochs of the thick and thin disc 
formation. In the \citet{Haywood2013} scenario, the thick disc ``formed from a well mixed interstellar
medium, probably first in starburst, then in a more quiescent mode, over a time scale of 4-5 Gyr''. They also
suggest that ``the youngest thick disk set the initial conditions from which the inner thin disk started to form 8 Gyr ago, 
at [Fe/H] in the range of (-0.1,+0.1) dex and \aabun=0.1 dex''. Finally, they interpret the low-metallicity
tail of the thin disc as having an outer disc origin, somewhat disconnected from the inner disc evolution.
In addition, similar separations in the \aabun ~versus [Fe/H] plane are distinguishable in the \citet{Ramirez} 
oxygen-abundance versus [Fe/H] measurements. These authors also report similar age-\aabun ~regimes as in \citet{Haywood2013}.

In summary, the increasing number of Galactic disc stars with known kinematical and chemical 
characteristics has emphasized the question of the continuity or not
between the thick and the thin disc components, in structural properties, kinematics, 
abundances and ages. Different authors like \citet{Norris99} and \citet{Bovy2012a} have even raised the possibility
of considering the Galactic disc as a single component, without a thick/thin disc distinction, analysing for instance 
the mass-weighted scale-height distribution of disc stars.

In this context, the GIRAFFE observations of the Gaia-ESO Survey \citep[GES,][]{GESMessenger} offer a unique opportunity 
to extend the previously high-resolution studies of the solar neighbourhood to larger and more radially extended samples.
In its first data release, GES already includes $\sim$10\,000 spectra (R$\sim$20\,000 and R$\sim$16\,000) of stars in the Milky Way field, 
from the halo, the thick disc, the thin disc and the bulge. The expected final number of targets, at the end of the
five years of observations, is $\sim$100\,000.

The present paper addresses the thick-thin disc transition as seen in the GES kinematical and chemical (\aabun ~and global
metallicity) first release data of FGK-type stars. Our point of view, exploiting the high resolution of the GES data, 
is that of the chemical characterisation and definition of the disc populations (essentially identified as thin and thick discs).
As a consequence, the subsequent analysis of the structural properties, kinematics and orbital parameters will be done
in the framework of the chemically defined disc populations.
Section~2 describes the data sample, including the stellar parameters and
\aabun~ measurements. Section~3 presents the derivation of the stellar distances and kinematics. Section~4 focuses on
the analysis of the number density distribution in the \aabun~ versus \met \ plane. Then, the distributions of the distances
to the Galactic plane (Section~5), of the azimuthal velocities (Section~6), the azimuthal and vertical velocity
dispersions (Section~7) and the stellar orbital parameters (Section~8)
are discussed, in the context of a proposed thick-thin disc separation based on the chemical criteria. Section~9 presents 
the rotational and abundance gradients with galactocentric radius and, finally, we discuss our results in Section~10.


\section{The GES data sample, the stellar parameters and the \aabun~ measurements.}
\label{Sect:Data}
The present work is based on the data collected by the Gaia-ESO Survey during the first nine months of observations.
As explained in \citet{GESMessenger}, the GES consortium has a work package structure that manages the data flows,
from target selection, through data reduction, spectrum analysis, astrophysical parameters determination, calibration
and homogenisation, to delivery of science data for verification analysis. The general data processing is described
in \citet[][in preparation]{DR1}.

\begin{figure}[ht]
\includegraphics[width=9cm,height=7cm]{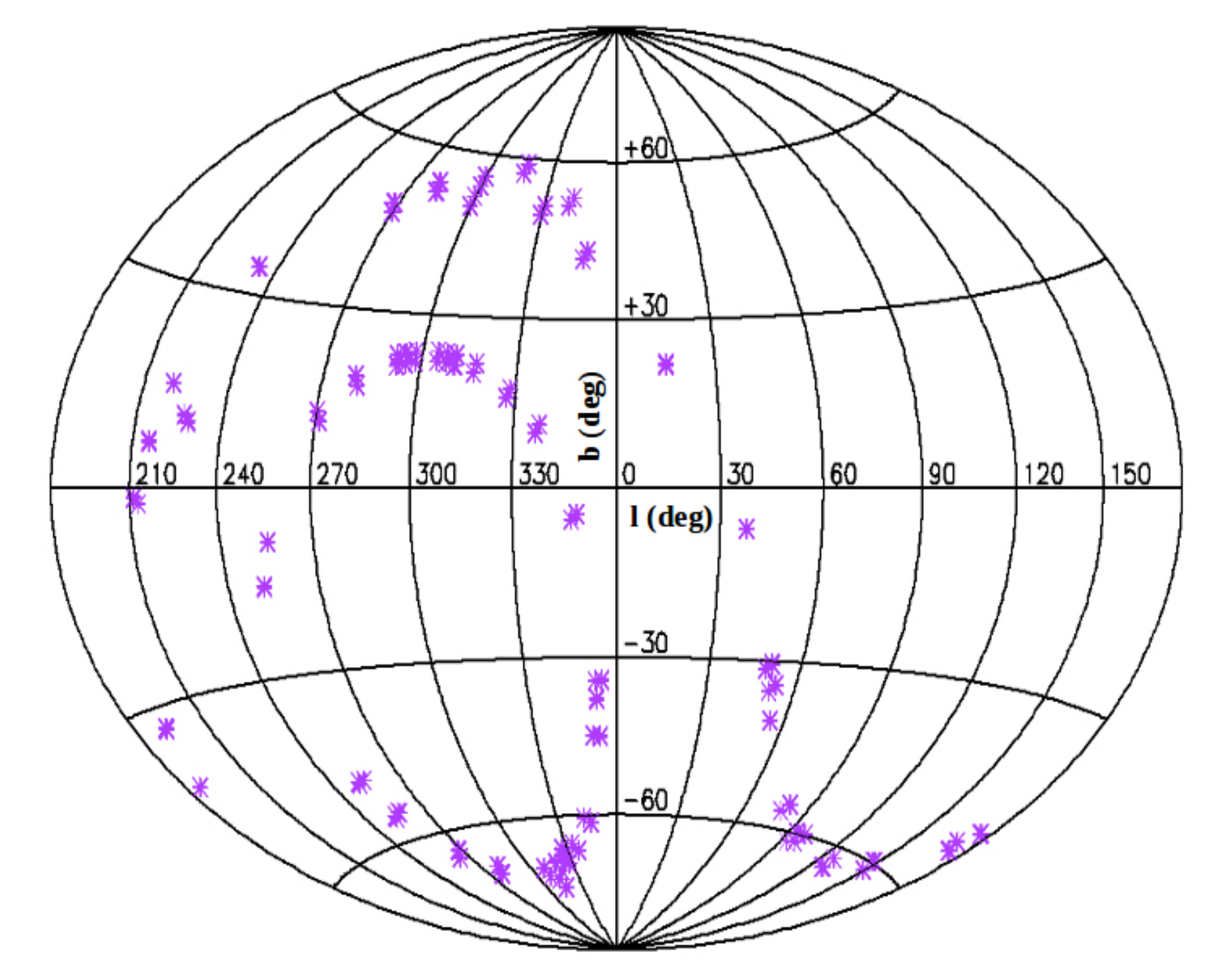} 
\caption{l and b coordinates for the lines of sight of the GES iDR1 (bulge and
star clusters fields are excluded).}
\label{lb}
\end{figure}

In this paper, we analyse the iDR1 GES results for $\sim$5\,000 Milky Way disc field stars observed with the GIRAFFE spectrograph:
4\,534 stars observed with both the HR10 (R$\sim$19\,800) and the HR21 (R$\sim$16\,200) setup modes, and 394 stars 
with only HR10 spectra. Fig.~\ref{lb} shows the l and b coordinates for the lines of sight of the GES iDR1 (bulge and
star clusters fields are excluded).
All the targets were selected from Visible and Infrared Survey Telescope for Astronomy (VISTA) photometry. 
As explained in  \citet{GESMessenger}, the target selection for the Halo and disc targets is based on the
Disc/Halo transition seen in the Sloan Digital Sky Survey photometry at 17 $< r <$ 18 and 0.2 $< g-r < $0.4. 
The equivalent selection from VISTA near-infrared photometry is used by GES with two main selection boxes:
\begin{itemize}
\item Blue box: 0.0$<$ J-K $<$0.45 and 14.0 $<$ J $<$ 17.5 mag
\item Red box: 0.4$<$ J-K $<$0.70 and 12.5 $<$ J $<$ 15.0 mag
\end{itemize}

The colour boxes are shifted according to the Schlegel extinction maps \citep[][]{schlegelMapas}. In addition, when 
the stellar density was not
enough to fill in all the FLAMES fibers, as for high latitude fields, the red box was extended to around J-K $<$0.85
and J$<$17 mag to allow for second priority fibers. 
The spectrum analysis has shown that the blue box has mainly selected dwarf stars with \teff ~between 6500~K and 5250~K. The corresponding
metallicity distribution peaks at around -0.6~dex for high latitude fields ($|b|>$30) and at -0.3~dex for low latitude fields. 
On the other hand, the red box has targeted dwarfs with \teff ~between 6000~K and 4500~K
and a small number of giant stars. The metallicity distribution of the red box targets is quite independent on the latitude and it 
peaks around -0.4~dex. Finally, the extention of
the selection boxes in the low density fields has added main sequence dwarfs with \teff ~between 6000~K and 4000~K with a doubled peaked
metallicity distribution (peaks at $\sim$-0.8~dex and -0.3~dex). The signal to noise value of this second priority targets is in 
general lower than the main targets data. As a consequence of the selection criteria, the thin disc is preferentially targeted in the
low latitude fields with respect to the thick disc, in a proportion of about 3:1, against about 1:1 for the high latitude fields.

The radial velocity measurements are based on a spectral fitting technique, that
includes a first guess determination of atmospheric parameters. The method is described in detail by \citet[][in preparation]{GESVrad}.
Typical errors in radial velocity are of the order of 0.3~km/s. The stellar atmospheric parameters and \aabun ~abundances were
determined by the GES work package in charge of the GIRAFFE spectrum analysis for FGK-type stars. The final recommended
effective temperatures (\teff), surface gravities (\logg), global metallicities (\met) and \aabun ~abundances (\aabun) 
were estimated thanks to the
combination of three different procedures: MATISSE \citep{Matisse}, FERRE \citep[][and further 
developments]{Ferre} and SME \citep[][and further developments]{SME}. The relative error distributions peak at 70~K for \teff, 0.10~dex for
\logg, 0.08~dex for \met ~and 0.03~dex for \aabun. The homogeneity of the
parameters between the stars observed with HR10 and HR21 and those observed with HR10 only was verified during
the GES parameters validation process. The adopted reference solar abundances are those of  \citet{Grevesse07} 
and the \citet{Gustafsson08} MARCS model atmospheres were used for the analysis.
For more details about the related GES parametrisation pipeline we refer the reader to \cite{WP10DR1}. 
In addition, 95\% of the targets have proper motions available from the PPMXL catalogue
\citep{ppmxl} with typical errors of 8~mas/yr.

\begin{figure}[ht]
\includegraphics[width=9cm,height=11cm]{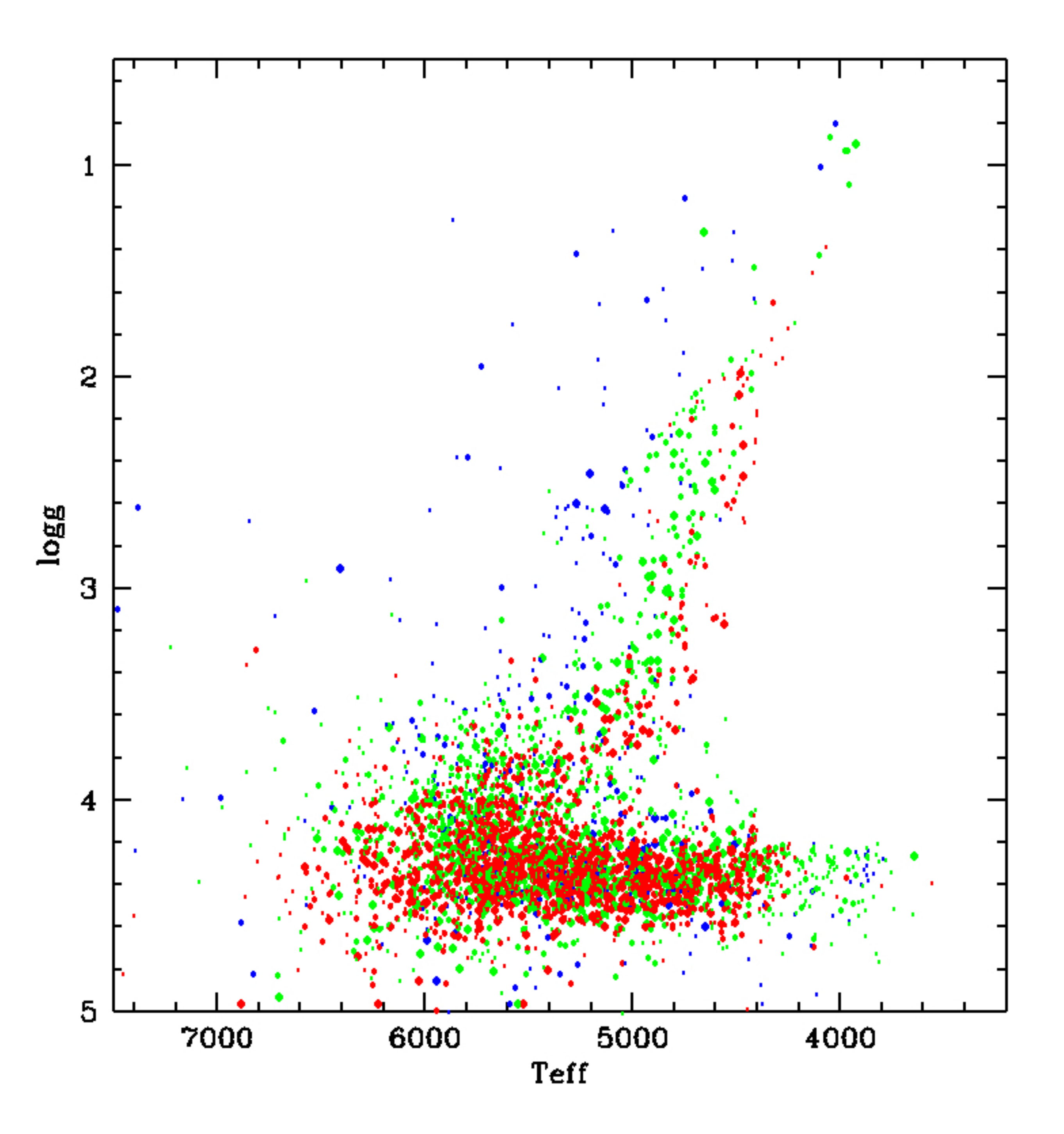} 
\caption{Hertzsprung-Russel diagram of the selected sample of GES stars. Red points correspond to stars with [M/H]$> -0.5$~dex, 
green points are intermediate-metallicity stars ($-1.0$~dex $< [M/H] \le $-0.5~dex), and blue points
are metal-poor objects ($[M/H] \le -1.0$~dex). To enhance the best quality data, the points size is inversely proportional 
to the associated errors in the atmospheric parameters.}
\label{MW_HRD}
\end{figure}

Fig.~\ref{MW_HRD} shows the Hertzsprung-Russel diagram of the analysed sample of 4\,928 GES stars, colour coded by metallicity.
Dwarfs stars represent about 74\% of the total sample.

\subsection{The global \aabun~ estimations}
The GES \aabun~ estimations used in this paper were determined, at the same time as the
stellar atmospheric parameters, by the three above mentioned groups with different methodologies
\citep{WP10DR1}.  More particularly, the final GES iDR1 \aabun~ values are the mean of the three independent determinations:
the FERRE and the MATISSE estimations, based on a four dimensional (\teff, \logg, \met, and \aabun) 
grid of synthetic spectra \citep[see][for a description of the methodologies adopted for this grid]{GrilleAMBRE}; 
and the SME determination, based on a mean of the preliminary individual
abundance estimations of Mg, Ca, Ti and Si. 
This global \aabun~ is estimated in the first analysis phase, together with \teff, \logg \ and \met~, and 
before the final GES recommended individual abundances are determined.

As explained in \citet{WP10DR1}, the final global \aabun~ 
reflects the information on the $\alpha$-element abundances present in the analysed wavelength
ranges for each type of star and spectrum quality. This approach, similar to the SDSS/SEGUE one
described by \citet{LeeAlphas}, estimates the overall $\alpha$-elements behaviour. 
The relation between the GES iDR1 global \aabun~  and the individual $\alpha$-element abundances,
also included in the GES iDR1 (Mg, Ca, Ti and Si), is presented in \citet{WP10DR1}. This analysis shows that the
global \aabun~ is dominated by the Mg abundance, with a standard deviation of the difference
equal to 0.069~dex.

Finally, GES iDR1 measurements of the Fe abundance are also available, although they are not used in this paper. 
The relation between \met~ and [Fe/H] has also been verified and it is presented in \citet{WP10DR1}.
The standard deviation of the difference between \met~ and [Fe/H] is 0.10~dex


\section{Distance estimations and kinematics}
\label{Sect:Dist}

The determination of the stellar distances is based on the method presented in \citet{Kordopatis11b} and successfully applied in \citet{Gazzano13} and \citet{Kordopatis13}.
This method projects the atmospheric parameters \teff, log$g$,  \met\ and their errors on a set of isochrones, obtaining the most likely absolute magnitude of the star, given an a priori knowledge on the lifetime spent by a star on each region of the HR diagram (i.e.  main-sequence stars are, statistically, more likely to be observed).

The isochrones on which the atmospheric parameters are projected are the ones of Yonsei-Yale \citep{Demarque04},  combined with the colour tables of \citet{Lejeune98}. Using the provided interpolation code, we generated a set of isochrones with a constant step in age of 1~Gyr, starting from 1~Gyr to 14~Gyr. As far as the metallicities are concerned, the isochrones are within a range of $-3.0 < \feh < 0.8$~dex, constantly spaced by 0.1~dex. 
The $\alpha-$enhancements of the isochrones have been selected in the following way, according to the typical \aabun~ of disc
and halo stars: 
\begin{itemize}
\item $\feh \geq -0.1$~dex, then \aabun$=0.0$~dex
\item $-0.1 \leq \feh \leq -0.3$~dex, then \aabun$=+0.1$~dex
\item $-0.4 \leq \feh \leq -0.6$~dex, then \aabun$=+0.2$~dex
\item $-0.7 \leq \feh \leq -0.9$~dex, then \aabun$=+0.3$~dex
\item $\feh \leq -1$~dex, then \aabun$=+0.4$~dex
\end{itemize}

On the other hand, \citet{Zwitter10}, in a similar approach to the one used here, shown that adopting isochrones of different alpha abundances does not affect the distance determination significantly.

In addition to the atmospheric parameters, the $(J-K_s)$ colours are also  used in order to obtain the most likely absolute magnitude of the star. Nevertheless, in order to do so, one must correct by the effect of the reddening, which is a function of the stellar distance and of the dust distribution. For that reason, we proceed in two steps. First, we compute the distances without using the colours, and we estimate iteratively the extinction at the distance of the star. Then, we apply the colour correction to the $(J-K_s)$ of the star, and re-compute the distances one final time using the de-reddened colour.

\subsection{Extinction correction}
The band in which we compute the distance modulus is the $J_{VHS}$ band. For the brightest  targets, no $VHS$ (VISTA Hemisphere Survey) photometry is available, due to the saturation limits of the survey, and hence we  adopt the 2MASS photometry,  in which system we have $J_{2MASS}=J_{VHS}$ and $K_{2MASS}=K_{VHS}$.  
In addition, the $(JK)_{ESO}$ provided by  the Yonsei-Yale isochrones match very well the $(JK_s)_{2MASS}$ and $(JK_s)_{VHS}$, so no colour transformation is needed when manipulating the magnitudes from the different photometric systems \citep{Carpenter01}. 

Once the absolute magnitude of the star is obtained using the \citet{Kordopatis11b} pipeline, a first estimation of the distance of the star is obtained,  assuming the \citet{Schlegel98} reddening.
For the lines-of-sight with the largest reddening ($E(B-V)$ higher than 0.10~mag), we have applied the correction suggested by \citet{Bonifacio00}:

\begin{equation}
 E(B-V)_a=0.10+0.65*[E(B-V)_{sch}-0.10], 
\end{equation}
where $E(B-V)_{sch}$ is the \citet{Schlegel98} reddening, and $E(B-V)_a$ is the adopted one. The extinction in the  $(JK_s)$  bands is then estimated using the following equations from \citet{McCall04}: 
\begin{eqnarray}
A_J=0.819*E(B-V)_a \\
A_{K_s}=0.350*E(B-V)_a.
\end{eqnarray} 

By applying the above extinction on the apparent magnitude and deriving the line-of-sight distance, we over-estimate the distance of the stars because the \citet{Schlegel98} reddening is a value integrated along the entire length of the line-of-sight.
We hence compute the following correction factor to apply for the Schlegel reddening, assuming an exponential disc of dust:   
\begin{equation}
E(B-V)_{corrected}=E(B-V)_a * (1-e^{\frac{-|D~sin(b)|}{h_{dust}}}),  
\end{equation} 
where $h_{dust}=125$~pc is the scale-height of the disc of dust causing the reddening \citep{Misiroitis}, $D$ is the 
line-of-sight distance and $b$ is the Galactic latitude of the target. 

\begin{figure}
\centering
$\begin{array}{c}
\includegraphics[width=1.0\linewidth]{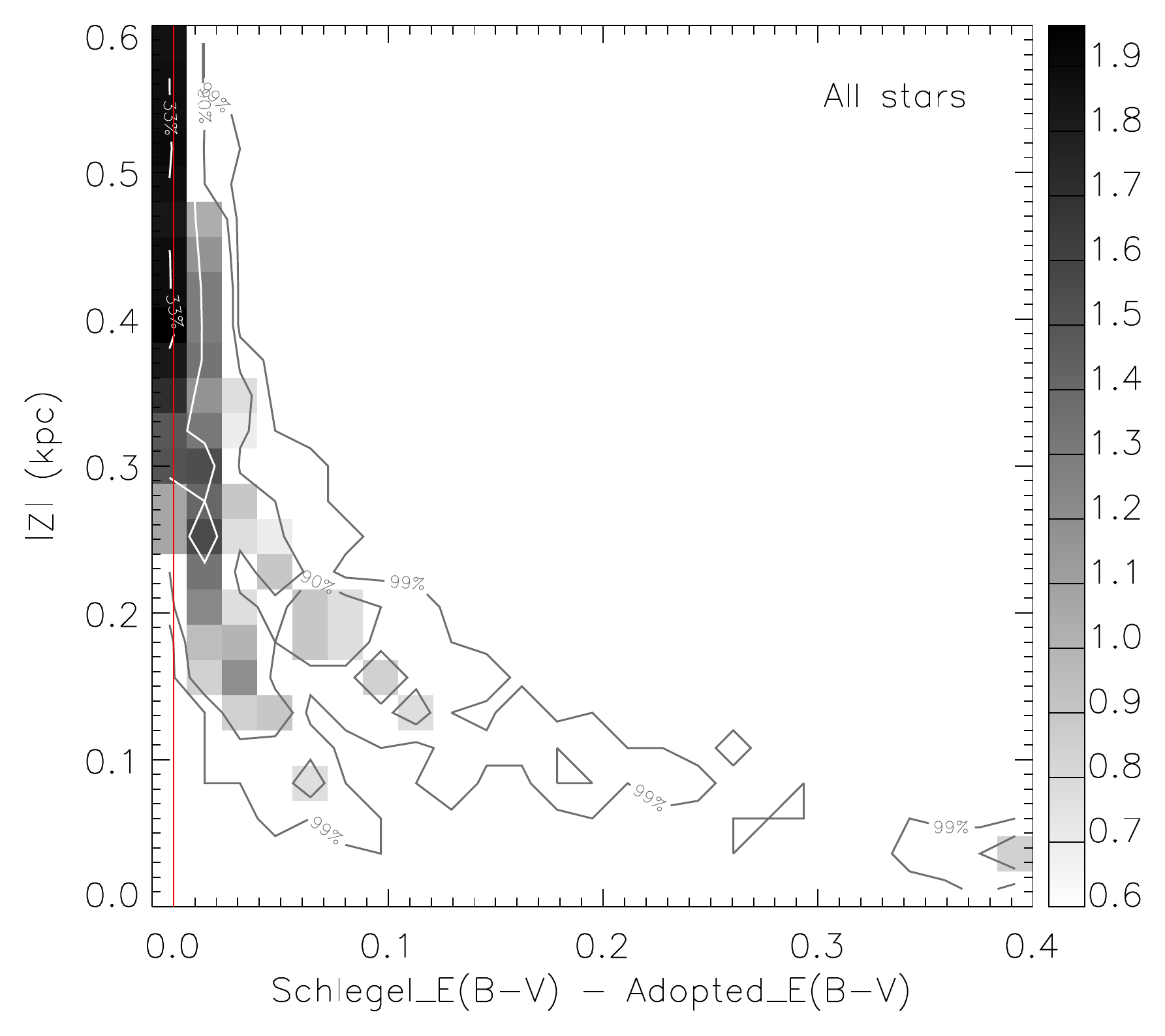}
\end{array}$
\caption{Reddening correction computed for each star depending on its line-of-sight distance, as a function of its distance from the Galactic plane $Z$. The gray-scale colours represent the star-counts in  $\log_{10}(N)$. The contour lines contain 33\%, 66\%, 90\% and 99\% of the stars. As expected the corrections made to $E(B-V)_a$ are only for the stars located closer than 400~pc from the Galactic plane. }
\label{fig:Reddening}
\end{figure}

Once the correction applied, we recompute the distance to the star, and following \citet{Ruchti11}, we repeat iteratively the above step until the difference between two consecutive distance estimations are smaller than 2\%. 
For the few stars that lie within the dust lane, no more than four iterations are needed until convergence. The adopted reddening correction, as a function of the distance from the Galactic plane is shown in Fig.~\ref{fig:Reddening}.

\begin{figure}
\centering
$\begin{array}{c}
\includegraphics[width=1.0\linewidth]{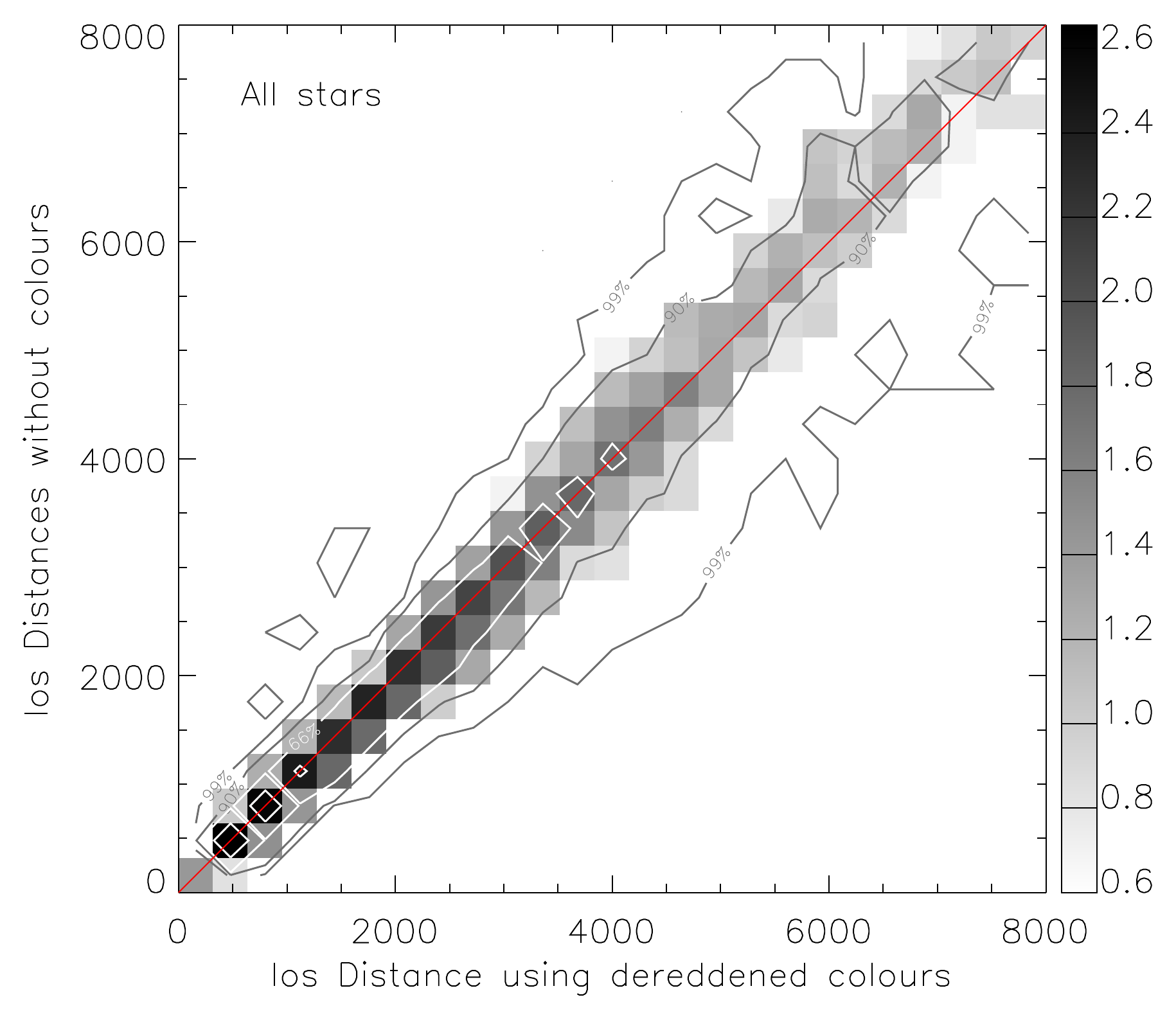}
\end{array}$
\caption{Change in the line-of-sight distance estimates when using only the effective temperature, surface gravity and iron abundance ($y-$axis) and when using the previously mentioned atmospheric parameters in combination with the $(J-K_s)$ colour ($x-$axis). The differences are mainly noticeable for the most  distant stars. The colour-coding and the contour-lines are the same as in Fig.~\ref{fig:Reddening}}
\label{fig:Distances_with_colours}
\end{figure}

Once the final extinctions have been computed, the de-reddened $(J-K_s)$ colour of the star is estimated, and the pipeline in order to estimate the absolute magnitude of the star is ran, using this time the $(J-K_s)$ colours.
The effect on the derived line-of-sight distances is plotted in Fig.~\ref{fig:Distances_with_colours}.  The latter distances are the ones that are used in the remainder (see for instance Fig.~\ref{RZ}).

\begin{table*}
\caption{Derived stellar distances(D), positions (in the R and Z coordinates) and kinematics (radial, azimuthal and vertical velocities), plus their associated errors ($\Delta D$,$\Delta R$,$\Delta Z$,$\Delta V_R$,$\Delta V_phi$ and $\Delta V_Z$).}
\centering
\begin{tabular}{c c c c c c c c c c c c c}
\hline\hline
ID &  D &$\Delta D$ & R & $\Delta R$ & Z & $\Delta Z$ & $V_R$ & $\Delta V_R$ & $V_phi$ & $\Delta V_phi$ & $V_Z$ & $\Delta V_Z$ \\
gir00184577-4700293    &   0.819    &   0.081    &   7.783    &   0.021    &  -0.766    &   0.076    &        17    &        37    &       253    &        37    &       -24    &        13   \\
gir00184695-4659371    &   2.113    &   0.113    &   7.451    &   0.028    &  -1.975    &   0.106    &       -12    &        97    &       203    &       100    &        25    &        37   \\
... & ... & ... & ... & ... & ... & ... & ... & ... & ... & ... & ... & ... \\
\hline
\end{tabular}
\label{distkin}
\end{table*}

\begin{figure}
\centering
$\begin{array}{c}
\includegraphics[width=1.0\linewidth]{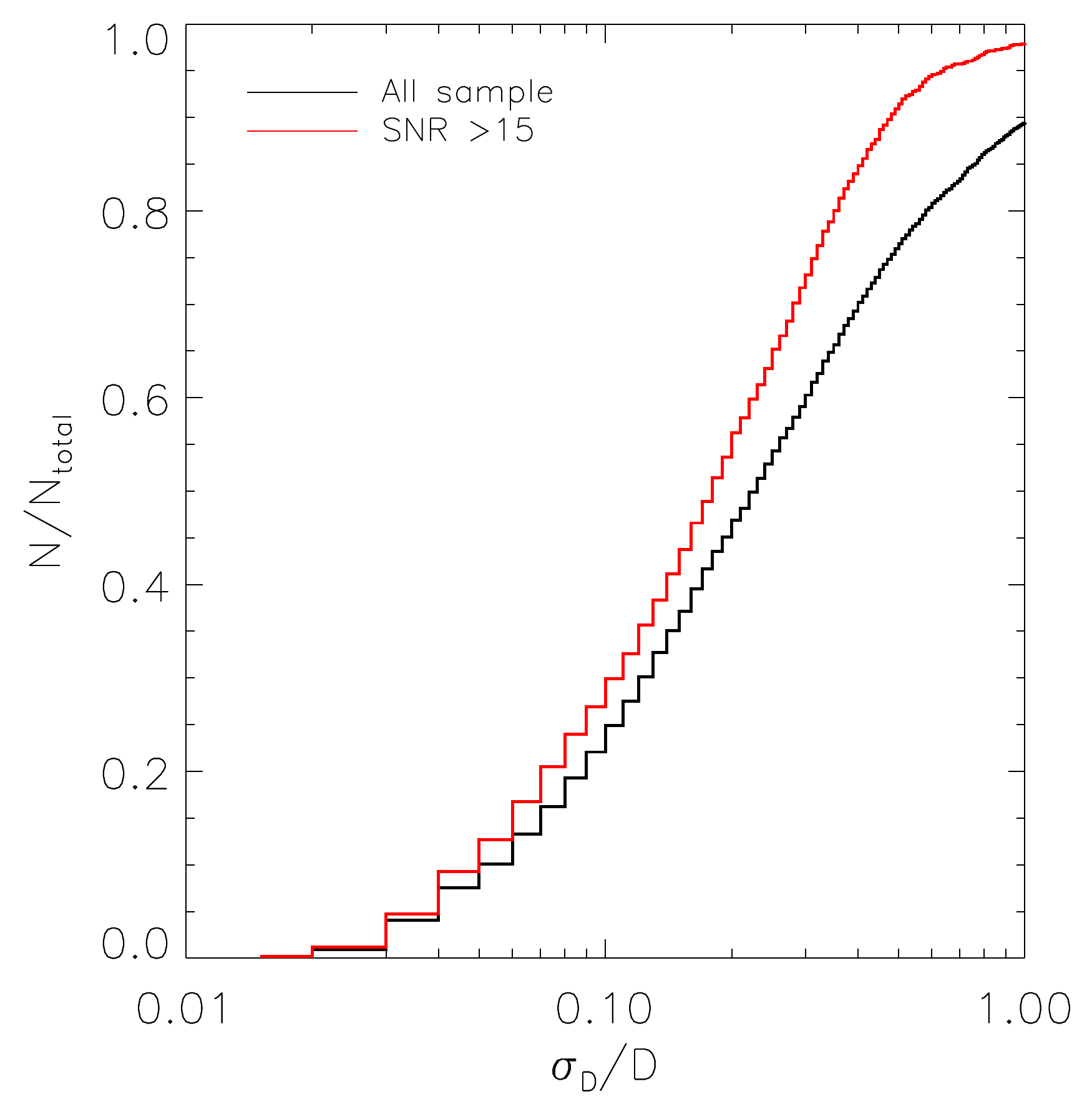}
\end{array}$
\caption{Cumulative histogram of the derived distance errors, for the entire sample (black line) and for stars with signal-to-noise ratios
better than 15 (red line).}
\label{errordist}
\end{figure}

\subsection{Errors in the line-of-sight distances}
In order to estimate the errors in the derived distances,  we performed 5\,000 Monte-Carlo realisations on the errors on 
the distance modulus. Those errors include the error on the derived absolute magnitude $J$  (which takes into account the errors in \teff, \logg, \met), and the error on the apparent $J$ and $K_s$ magnitudes.  The uncertainty in the atmospheric parameters is taken into account when computing the absolute magnitude, in the same fashion as described in \citet{Kordopatis11b}, i.e. by estimating the dispersion of the weighted absolute magnitude of the stars on the isochrones.  
The parameter dominating the error in the distances is \teff \ for main sequence stars and \logg \ for giants.

\begin{figure}
\centering
$\begin{array}{c}
\includegraphics[width=1.0\linewidth]{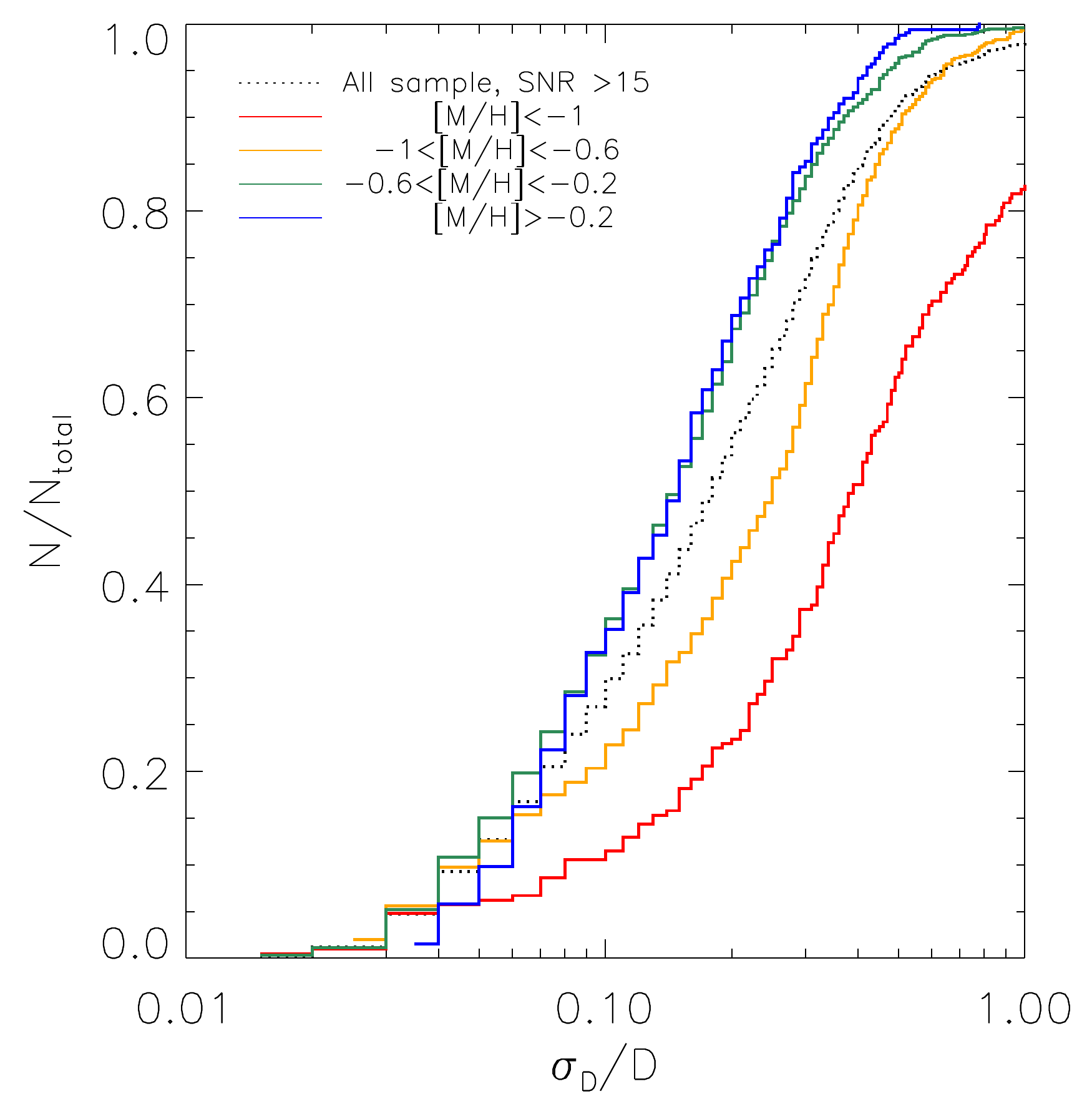}
\end{array}$
\caption{Cumulative histogram of the derived distance errors for different metallicity bins, for stars with SNR$>$15.}
\label{errordistMet}
\end{figure}

\begin{figure}
\centering
$\begin{array}{c}
\includegraphics[width=1.0\linewidth]{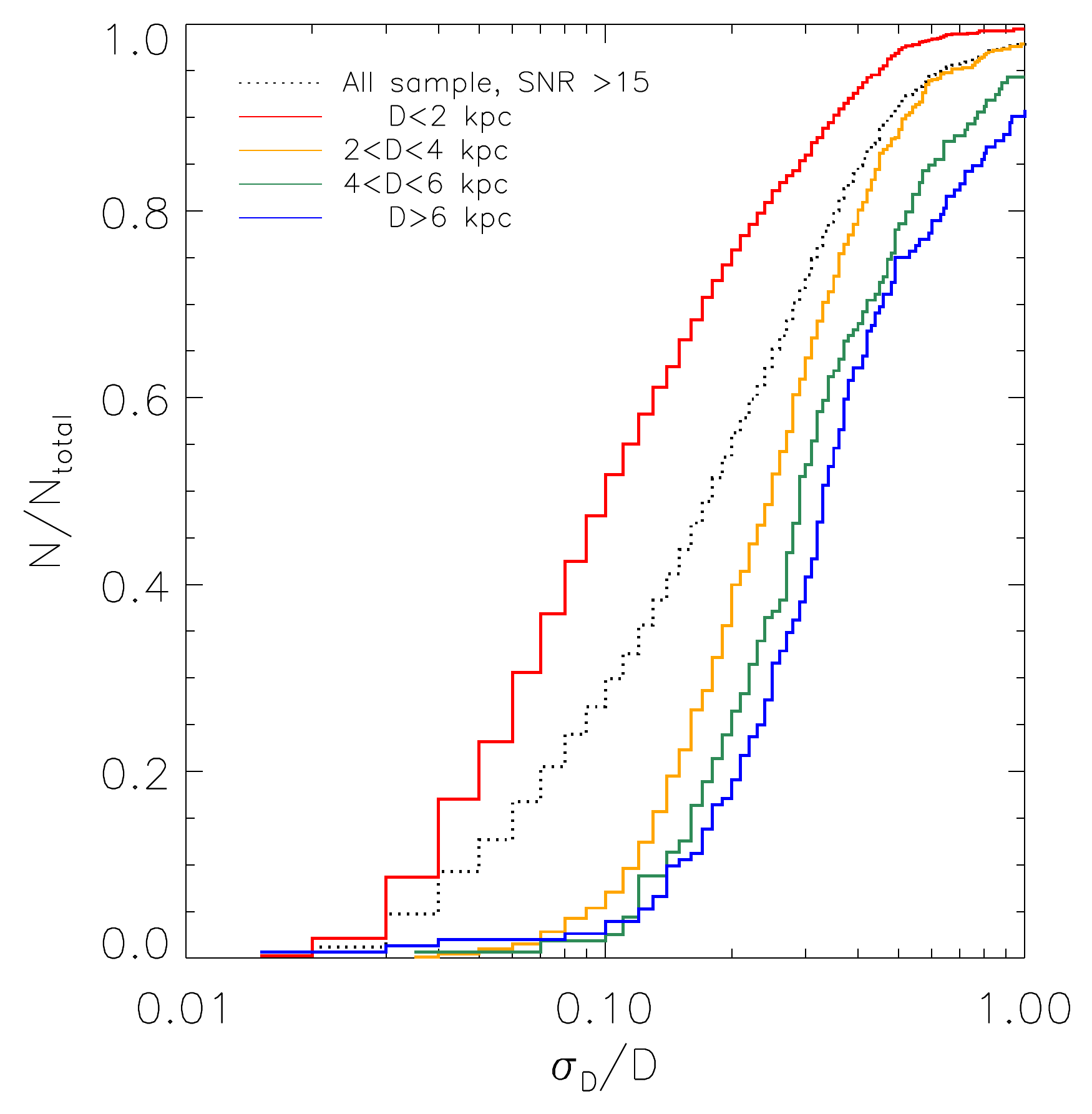}
\end{array}$
\caption{Cumulative histogram of the derived distance errors for different distance bins, for stars with SNR$>$15.}
\label{errordistDist}
\end{figure}

Fig.~\ref{errordist} shows the cumulative histogram of the derived distance errors, for the entire sample and for a 
sub-sample with signal-to-noise ratios higher than 15 and errors in \teff \ and \logg \ lower than 400~K and 0.5~dex, 
respectively. $\sigma_D$ is the inferred error bar when the above factors are taken into account. 
As shown in Fig.~\ref{errordist}, 45$\%$ of the total sample has $\sigma_D/D<20\%$ and 80$\%$
has $\sigma_D/D<50\%$. These values increase to 55$\%$ and 90$\%$ for the selected sub-sample of SNR$>$15 (2107 stars).
In addition, Fig.~\ref{errordistMet} and Fig.~\ref{errordistDist} show the dependences of the distance errors on the metallicity
and the distance value itself, for stars with SNR$>$15.
The final errors in positions  for the individual stars are included in Table~\ref{distkin}. 

\subsection{Validation of the distances}

\begin{table*}
\caption{Adopted globular cluster parameters \citep[from][in the catalogue updated version of December 2010]{Harris96}.
The two last columns show the derived distances with their associated errors (quadratic sum of the standard deviation 
of the members distances and  their median individual errors).}
\centering
\begin{tabular}{c c c c c c c}

\hline\hline
Cluster & los distance & $E(B-V)$ & $\feh$ & \vrad & $\sigma$\vrad & derived los distance \\
& (kpc) & (mag) & (dex) & (\kms) & (\kms) & (kpc)\\ 
\hline

NGC 5\,927 & 7.7  & 0.45 & -0.49 & -107.5 & &  8.9 $\pm$ 3.3 \\
NGC 1\,851 & 12.1 & 0.02 & -1.18 & 320.5 & 10.4 & 11.3 $\pm$ 4.1 \\
NGC 2\,808 &9.6 & 0.22 & -1.14 & 101.6 & 13.4 & 9.4 $\pm$ 3.1\\
NGC 4\,372 &5.8 & 0.39 & -2.17 & 72.3 & & 8.1 $\pm$ 3.1 \\

\hline
\end{tabular}
\label{table:gc_values}
\end{table*}

We verified the accuracy of our results by evaluating the distances at which were found the globular clusters (GC) stars, observed and labelled as such by GES. The reference distances are the ones from \citet{Harris96}, in the updated version (December 2010) of the catalogue\footnote{http://www.physics.mcmaster.ca/~harris/mwgc.dat}, summarised in Table~\ref{table:gc_values}. In order to remove foreground contamination, in addition to using the GES labels, we have rejected all the targets that were outside the $3\sigma$ of the velocity dispersion  of the GC (an arbitrary value of $\sigma=$10~\kms was adopted when this information was not available) and had iron abundances lower or greater than $0.5$~dex of the mean cluster value (see Table~\ref{table:gc_values}).  

\begin{figure}
\centering
$\begin{array}{c}
\includegraphics[width=1.0\linewidth]{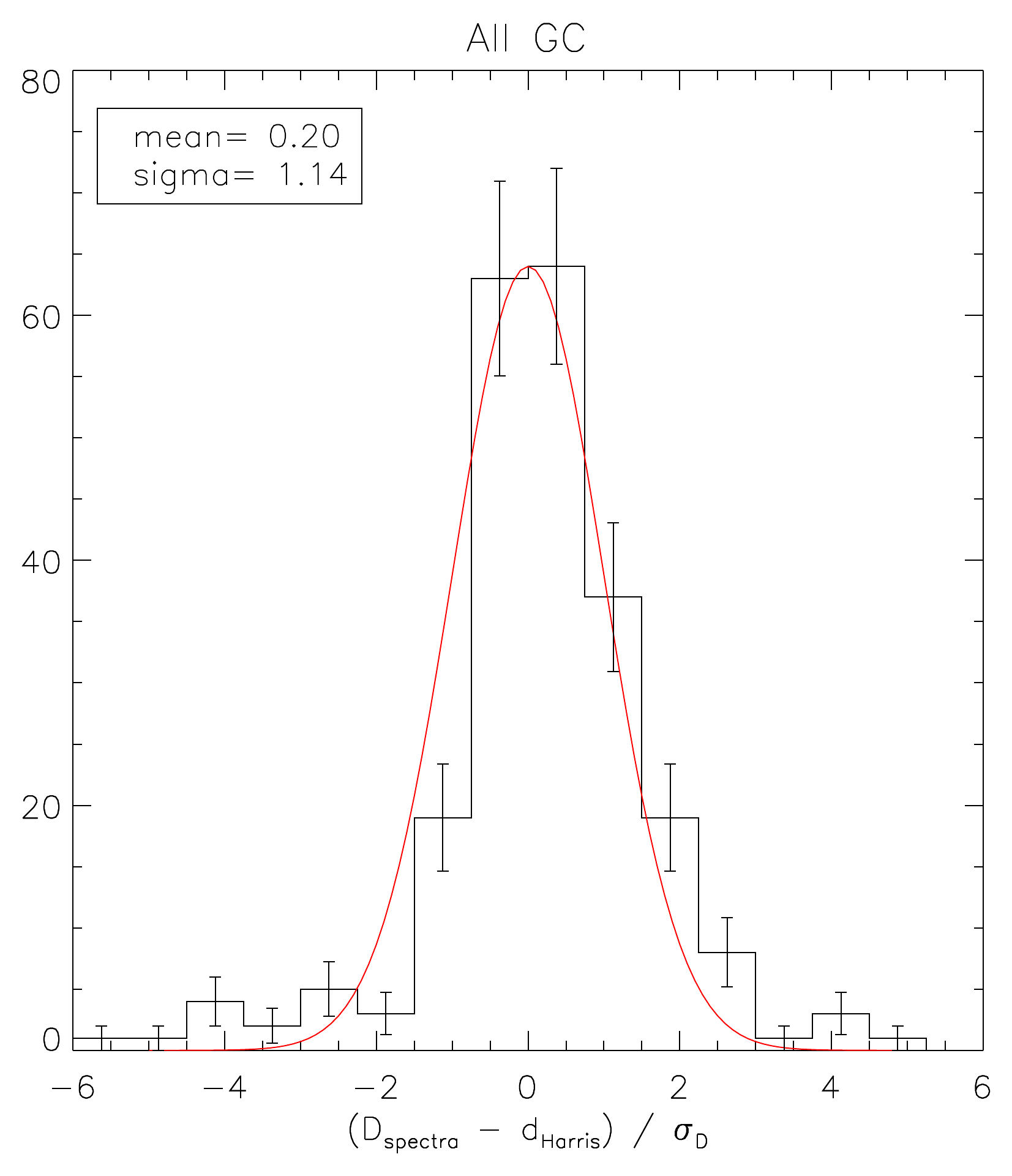}
\end{array}$
\caption{Histogram of the error distribution for all the stars selected as belonging to the globular clusters observed by GES. The red Gaussian is not a fit of the histogram but represents the unit  Gaussian which should follow perfect bias-free measurements with correct error estimations. A very satisfactory result is obtained, with an estimated bias of $\sim$5\% and a standard deviation of $\sim 14\%$.}
\label{fig:Distance_accuracy}
\end{figure}

Figure~\ref{fig:Distance_accuracy} shows the histogram of the difference between the derived distance, $D$ and the reference one, $d_{Harris}$, normalised by the estimated individual error on the distances $\sigma_D$ (obtained from our pipeline). The error bars indicate the Poisson noise, i.e. the square root of the stars for each bin. Three stars were removed from the figure, as they were lying respectively at $(D-d_{Harris})/\sigma_D=-581$,     $-293$ and    $-289$ and were considered as contaminators. 
In the case where our analysis is bias free and the errors correctly estimated, the histogram of Fig.~\ref{fig:Distance_accuracy} is supposed to be Gaussian centred on zero and of unit dispersion. We can see that it is the case, validating in that way our approach.  More precisely, once the three contaminators have been removed, we find a distribution centered at 0.20 and a standard deviation of 1.14. The mean value of the  $(D-d_{Harris})/\sigma_D$ distribution corresponds to a bias in the estimated distances 
($(D-d_{Harris})/D$) of less than $5\%$.
The value of the sigma indicates an under-estimation of the errors of $\sim 14\%$ with respect to the differences between the literature values and
the ones derived in this work. However, we would like to stress that the Harris’ errors on the distance of the GCs could probably explain most of
this difference.  On the other hand, the slight asymmetry of the histogram comes from the measurements of NGC~4\,372, which has also the lowest metallicity and hence the atmospheric parameters with the largest uncertainties.
In addition, it is worth noticing that the GES iDR1 globular cluster targets are giant stars, while the majority of the observed
field stars are main sequence objects. Although GES is a magnitude limited sample, 80$\%$ of the targets are placed at
smaller distances than the above analysed GCs.

Finally, it is worth mentioning that for dwarf stars, as explained in \cite{WP10DR1}, a degeneracy between \teff \ and
\logg \ appears progressively, as \teff \ decreases, 
starting around \teff$\sim$5000~K. This degeneracy gives rise to a negative bias (stars may appear cooler and at lower gravities). 
We have explored the consequence on the derived distances on the above mentioned degeneracy, taking into account that 
the distances are calculated after the projection of the atmospheric parameters on the isochrones. In the corresponding 
cool part of the main sequence, \teff \ is the main parameter influencing the absolute 
magnitude determination, as the isochrones points of different ages and metallicities are all tightly located in the HR diagram. 
In addition, as the slope of the isochrones is very low, the determined distances are not very sensitive to errors in \teff. 
Moreover, the projection of the
determined parameters on the isochrones, for the distance determination, corrects both \teff \ and \logg. This correction is
of the same order of the bias for stars with \teff$>$4500~K. For cooler stars (4500$>$\teff$>$4200~K), the residual bias 
in \teff, after the isochrone projection, is around 100~K.  This error in \teff implies around 6$-$7$\%$ of distance error, depending on the metallicity. 
For this work, the selected stellar subsample of iDR1 (c.f. Section~4) contains 
only 3$\%$ of stars cooler than 4500~K in the thin disc sequence, and 2$\%$ in the thick disc one. Therefore, in the worst case,
a bias of only $\sim$7$\%$ in distance, well inside the typical errors of the sample, can appear for about 2-3$\%$ of our final sample.

\subsection{Derivation of Galactocentric velocities}

From the derived line-of-sight distances, the three-dimensional Galactic positions are obtained for all the analysed stars
\citep[assuming $R_\odot=8$~kpc and $Z_\odot=0$~pc, see][]{Reid93}. The derived values together with their associated errors 
are reported in Table~\ref{distkin}.
Once the three-dimensional Galactic positions are obtained, the galactocentric radial, azimuthal, and vertical velocities are computed using the PPMXL proper motions, and the equations from the appendix of \citet{Williams13}.
We have used for $(U_\odot,V_\odot, W_\odot)=(11.1, 12.24, 7.25)$~\kms\ \citep*{Schonrich10} and the Local Standard of Rest (LSR) is at $V_{LSR}=220$~\kms. The obtained galactocentric velocities are reported in Table~\ref{distkin}.

\subsection{Error estimation in velocities}
In order to evaluate the errors on the stellar velocities, we performed 5\,000 Monte-Carlo realisations, considering that the proper motions, the radial velocities, and the line-of-sight distances are independent. 
Depending on the galactic coordinates (l,b) of a star,  the uncertainties on the distances, the proper motions and the radial velocities will affect the 3D-velocity estimations in a different manner. Typically, the dominant source for the uncertainty comes from the errors in proper motions (typical errors of 8 mas/yr, see Sect.2), whereas the errors on the radial velocities  (typically of 0.3 km/s) have in general a negligible effect.  As an example, for a given star at 1~kpc in 290$<$l$<$340 and 10$<$b$<$30,  the contribution to the 3D-velocity uncertainties will be of
5$-$10~km/s   for an error of 10$-$20\% on the distance and of 15$-$30 km/s  for an error of 5$-$10 mas/yr on the proper motions.
The final errors in  kinematics for the individual stars are included in Table~\ref{distkin}. 

\section{Chemical characterisation of the thin and the thick discs}

\begin{figure*}[ht!]
\centering
$\begin{array}{c}
\includegraphics[width=15cm,height=15cm]{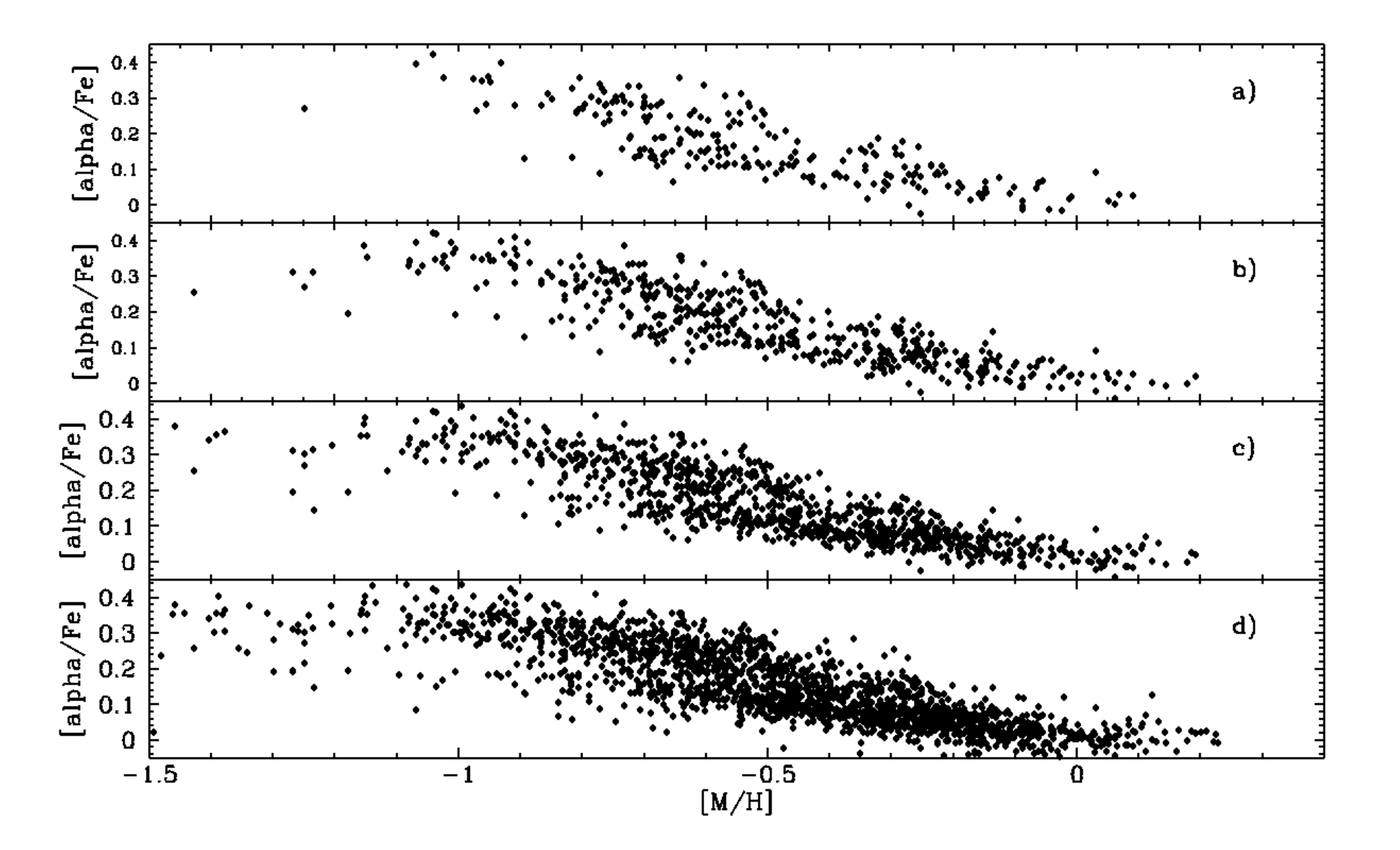}  
\end{array}$
\caption{$\alpha$-elements over iron abundance as a function of metallicity for four different sub-samples of stars with increasing
errors in the abundance determination. Panel a) shows the results for stars with errors in \met~ and \aabun~ smaller than 
0.07~dex and 0.03~dex, respectively (209 stars). Panel b) shows the results with errors smaller than 0.09~dex and 0.04~dex 
in \met~ and \aabun\ (505 stars). Panel c) illustrates the values for 1\,008 stars with errors smaller than 0.15~dex and
0.05~dex, respectively. Finally, panel d) shows all the stars with errors in \teff\ lower than 400~K, errors in \logg\ lower
than 0.5~dex and a spectral signal-to-noise ratio higher than 15 for the HR10 configuration (1\,952 stars).}
\label{AlfaFeMet}
\end{figure*}

\begin{figure*}[ht!]
\centering
$\begin{array}{c}
\includegraphics[width=15cm,height=15cm]{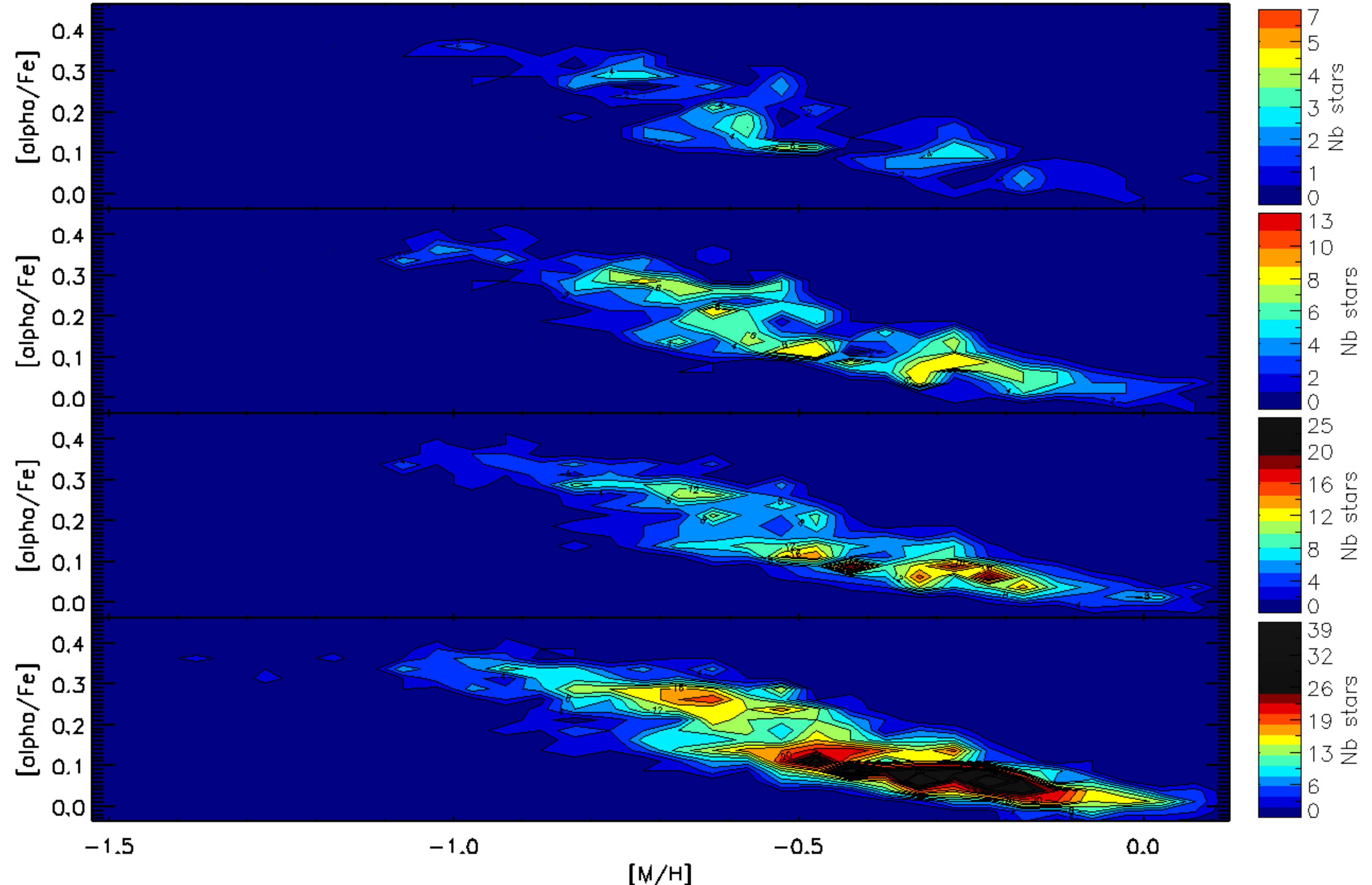}  
\end{array}$
\caption{Contour plots of the \aabun~ versus \met~ distributions of Fig.~\ref{AlfaFeMet}}
\label{AlfaFeMetContour}
\end{figure*}

Thanks to their high-resolution, the GES spectra offer an exceptional opportunity to characterise the distribution of disc stars
in the \aabun~ versus \met~ plane, with a robust enough statistical analysis and rather low abundance errors.
To this purpose, we have first selected sub-samples of stars with progressively higher errors in the abundance determination.
This allows, on one hand, to find the balance between number statistics and abundance 
accuracy, and on the other hand, to test the effects that low numbers of stars and high measurement errors could have
in the characterisation of the stellar populations.

 Fig.~\ref{AlfaFeMet}, shows the distribution of stars in the \aabun~ versus \met~ plane for
four progressively larger sub-samples with increasingly  higher error limits in the abundance determination.
Table~\ref{Samples} shows the maximum errors and signal-to-noise limits considered for each of the
analysed sub-samples. In addition, Fig.~\ref{AlfaFeMetContour} presents the same \aabun~ versus \met~ distributions
in a contour levels form, for a complementary illustration.

\begin{figure*}[ht!]
\centering
\includegraphics[width=17cm,height=12cm]{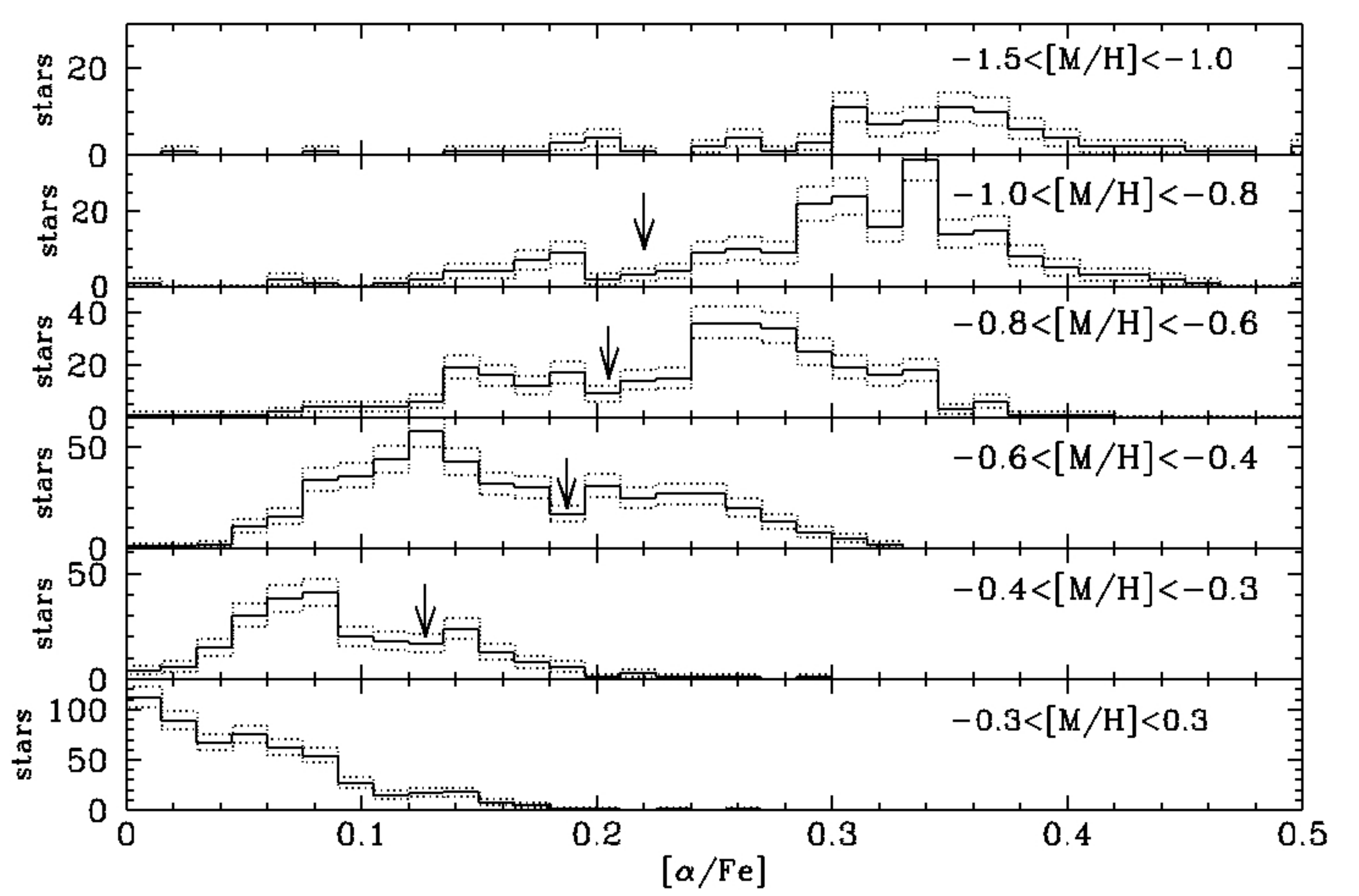}  
\caption{Distribution of \aabun~ values in six intervals of metallicity for the selected sub-sample (d) of Fig.~\ref{AlfaFeMet}.  Error bars, corresponding to Poisson uncertainties, are given by the dotted lines. The position of the low density region separating the thick and the thin disc sequences is marked with an arrow.}
\label{Gap}
\end{figure*}

The well-known decreasing evolution of \aabun~ with metallicity is present in all the panels, within the expected
values for disc stars. In addition, two sequences separated by a low density region can be identified: the
thick disc one in the high-$\alpha$ regime and the thin disc one in the low-$\alpha$ regime (see below for the selection criteria). 
Both sequences overlap
in metallicity from about -0.8~dex to -0.3~dex. For the last sub-sample of 1\,952 objects, the increase in the scatter 
of the measurements blurs the chemical separation, although not completely. 

\begin{table*}
\caption{Maximum errors allowed in the different considered sub-samples of stars.}
\centering
\begin{tabular}{c | c | c | c | c | c | c}
\hline\hline
Sub-sample id & Max error \teff & Max error \logg & Max error \met & Max error \aabun & HR10 SNR limit & Number of stars\\
 & (K) & (dex) & (dex) & (dex) & \\
\hline
a & 400~K & 0.50 & 0.07 & 0.03~dex & & 209\\
b & 400~K & 0.50 & 0.09 & 0.04~dex & & 505\\
c & 400~K & 0.50 & 0.15 & 0.05~dex & & 1\,008\\
d & 400~K & 0.50 &   $-$   &    $-$       & 15 & 1\,952\\
e & 400~K & 0.50 &  0.15 (\met$>$-1)      &  0.05 (\met$>$-1)        & 15   & 1\,016\\
  &       &      & 0.20 (\met$<$-1)  &  0.08 (\met$<$-1)  &  15  & \\
\hline\hline
\end{tabular}
\label{Samples}
\end{table*}

First, as shown in Fig.~\ref{Gap}, we have studied the position
of the low density region along the \aabun~ vs. \met~ plane, thanks to the analysis of the \aabun~
distribution in four intervals of metallicity. The largest sub-sample of stars in  Fig.~\ref{Gap}
(panel d of Fig.~\ref{AlfaFeMet}) was used for this purpose. The identified position of the gap in the \aabun~ vs. \met~ plane
will be used in the following as the separation criterion between the thick and thin disc populations for further analysis.
This low density region is found, after a visual examination of the histograms, at \aabun$=$0.13, 0.19, 0.21 and 0.22~dex for the four central considered \met\ intervals in Fig.~\ref{Gap}.

\begin{figure*}[th!]
\centering
\includegraphics[width=15cm,height=8cm]{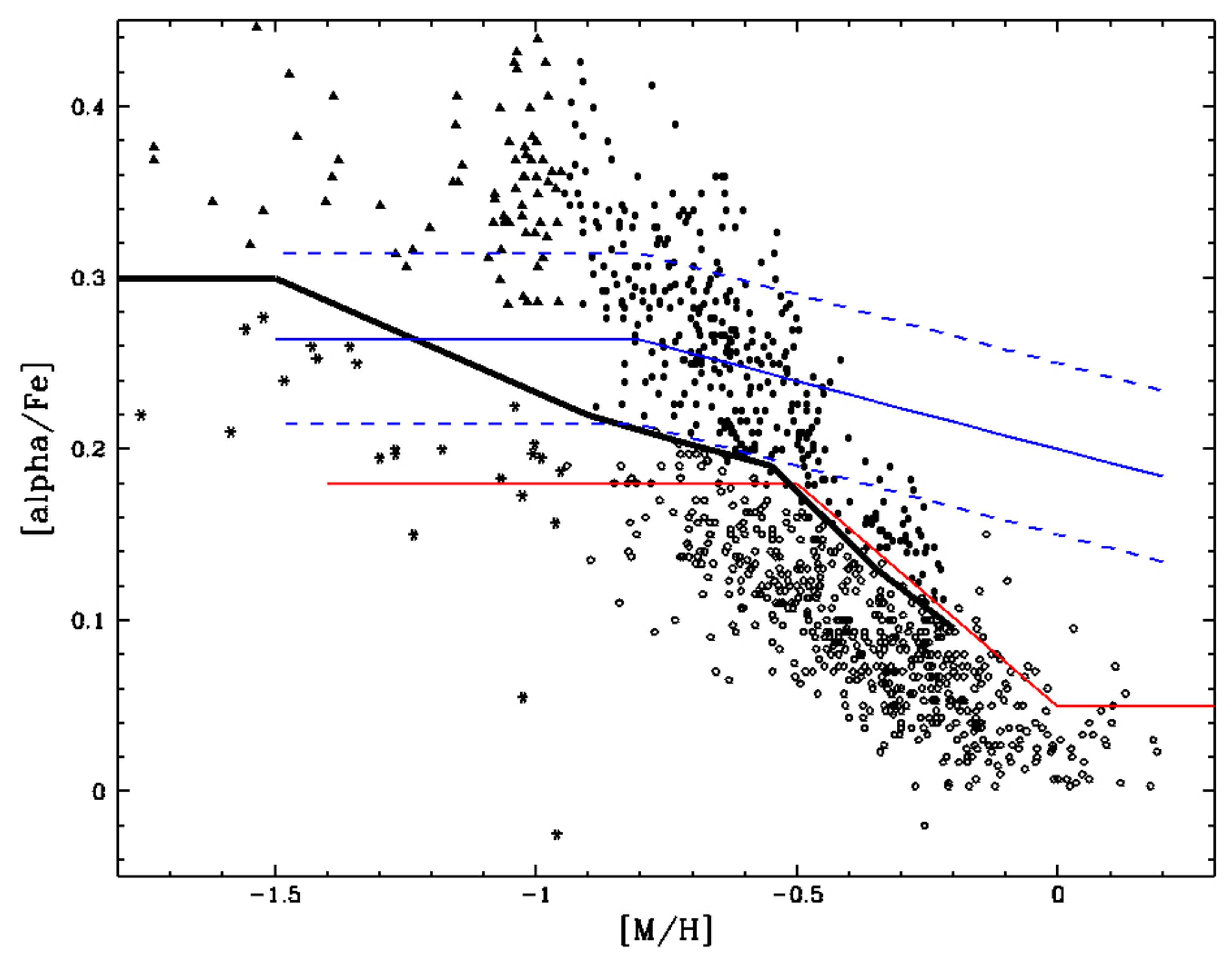}  
\caption{$\alpha$-elements over iron abundance as a function of metallicity for sub-sample {\it e}. Filled dots are thick disc sequence
stars, open circles are thin disc sequence objects. Metal-poor stars (\met$<-1.0$) are marked with filled triangles and asterisks,
depending if their \aabun \ are in the high-$\alpha$ or the low-$\alpha$ regime. The black line shows the proposed division between
the thin and thick disc sequences in agreement with Fig.~\ref{Gap}. The red line shows the separation proposed by high resolution
analysis of \citet{Adibekyan2012} for the solar neighbourhood. The blue solid line is the separation proposed by \citet{Lee2011b} from
low resolution SEGUE data and the two dashed lines indicate their dividing points for the thin and the thick disc stars to avoid the 
misclassification of stars.}
\label{Pops}
\end{figure*}

\begin{figure}[ht]
\includegraphics[width=9cm,height=11cm]{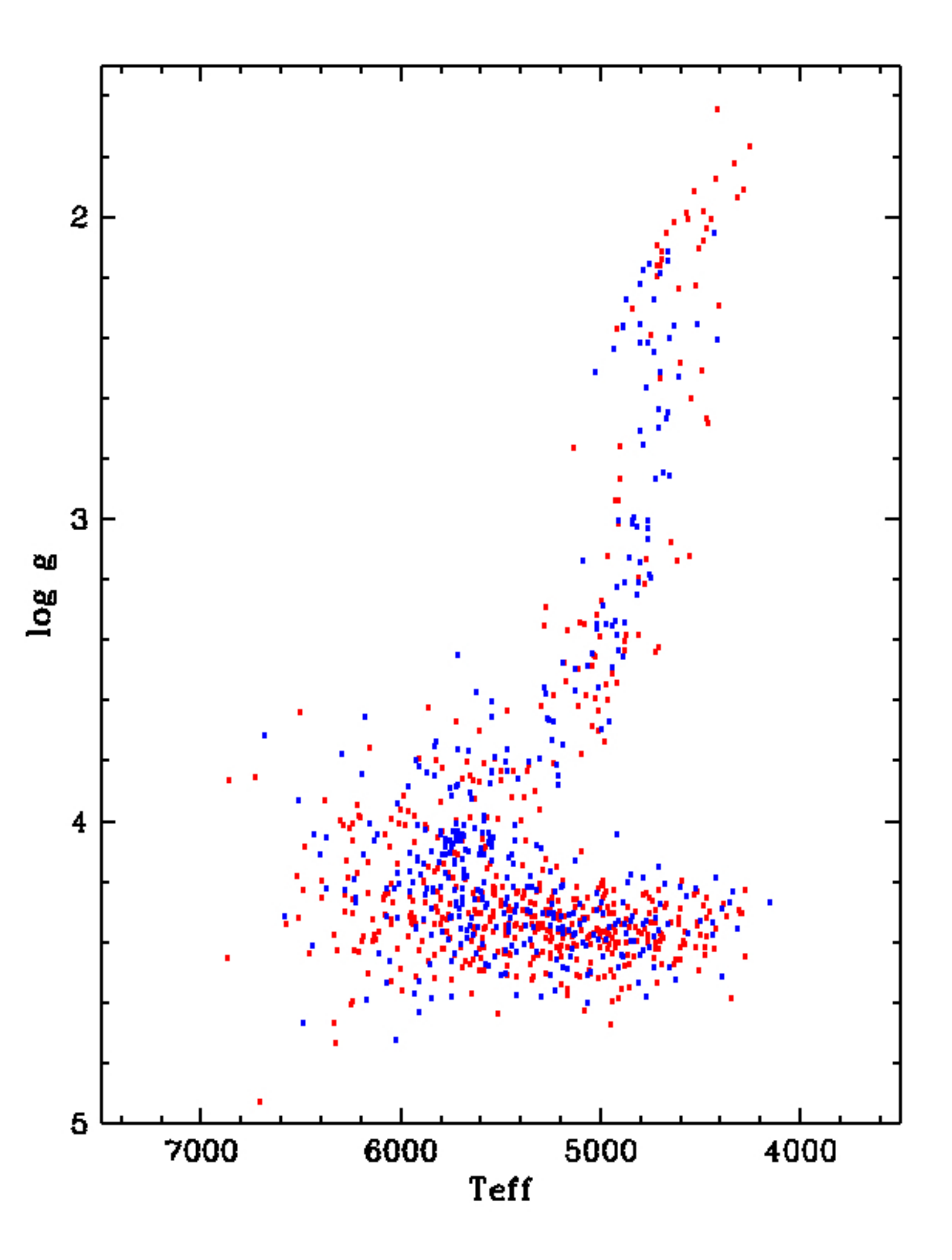} 
\caption{Hertzsprung-Russel diagram of the selected sub-sample {\it e}. Red and blue points correspond to stars in the 
defined thin and thick disc sequences, respectively.}
\label{HRD}
\end{figure}

Then, in order to more precisely characterise the thick and thin disc populations previously defined,
and to take into account of the natural increase of the abundance measurement errors for metal-poor star's spectra \citep{WP10DR1},
we have defined a new sub-sample of objects, hereafter called {\it e} (c.f. Table~\ref{Samples}). This new sub-sample contains all stars
with errors smaller than 0.15~dex and 0.05~dex in \met~ and \aabun, respectively for \met$>$-1~dex
plus metal-poor stars (\met$<$-1), with allowed errors as large as 0.2~dex in \met~ and 0.08~dex.
The final sample is shown in Fig.~\ref{Pops} and contains 1\,016 stars.  A black line (based on the
low density regions defined in Fig.~\ref{Gap}) is plotted in Fig.~\ref{Pops} to identify the separation 
between the thick and thin disc sequences.
Fig.~\ref{HRD} shows the Hertzsprung-Russel (HR) diagram of this selected sub-sample {\it e}. The red and blue points correspond 
to stars in the defined thin and thick disc sequences, respectively.

The thick disc sequence (filled circles in Fig.~\ref{Pops}) seems to extend from \met~$=-1.0~$dex and
\aabun~$=0.35$~dex to about \met~$=-0.25~$dex and \aabun~$=~0.1$dex, decreasing linearly with a slope of about -0.30.  
We have also considered metallicity bins of 0.2~dex and calculated the dispersion in \aabun~ for
each bin. The resulting mean \aabun\ dispersion is 0.042~dex. This can be compared to the mean error in \aabun\ 
that is equal to 0.03~dex. On the other hand, the thin disc sequence (open circles), seems to be present 
from about \met~$=-0.8~$dex and \aabun~$=~0.18~$dex, to about \met~$=0.2~$dex and \aabun~$=~0.0~$dex, 
decreasing with a slightly milder slope of about -0.20.

In addition, a high-$\alpha$ metal-poor population is 
also visible in our data, well above the Galactic plane (see Sect. 5), and occupying a flat sequence 
from \met$\sim$-1.0~dex down to the lowest sampled metallicities (filled triangles in Fig.~\ref{Pops}). 
The mean \aabun \ value of these halo stars is around 0.35~dex. This high-$\alpha$ metal-poor sequence 
could be identified with the classical high-$\alpha$ halo sequence or, at least partially, with the metal-weak
thick disc population \citep[e.g.][]{Ruchti11,Kordo13MWTD}.
Moreover, in the same metal-poor metallicity regime (\met~values lower than $\sim$ -0.9~dex), a metal-poor low-$\alpha$ sequence 
seems to exist (asterisks in Fig.~\ref{Pops}). 
After an individual inspection of these low-$\alpha$ metal-poor stars spectra, most of them 
have signal-to-noise values higher than 25 in the HR10 setup, and have thus reliable \met\ and \aabun\ measurements. 
When only the lower error measurements are 
considered (stars with errors smaller than 0.05~dex in \aabun~ and 0.15~dex in \met~ and signal-to-noise ratio for the HR10 spectra higher than 20), the low-$\alpha$ metal-poor stars represent about 10\% of the observed stars with \met~values lower 
than $\sim$ -0.9~dex. These stars could confirm the 
low-$\alpha$ metal-poor halo stars first observed by  \citet{NissenSchuster2010}, and also reported by \citet{Nissen11}, \citet{Schuster12},
 \citet{Adibekyan2012} and \citet{Ishigaki}. On the other hand, the possible link between those objects and the metal-weak thick disc population
\citep[e.g.][]{Morrison, Carollo, Kordopatis13} seems to be ruled out by their low \aabun \ ratios, as metal-weak thick disc stars are reported to
have \aabun$\sim0.3$~dex \citep{Ruchti11}. 
As a complementary description of these stars, the individual element abundances of these stars are analysed in \citet[][in preparation]{Sarunas}. 

\subsection{Comparison with literature studies}


\begin{figure}[th!]
\includegraphics[width=8.5cm,height=8cm]{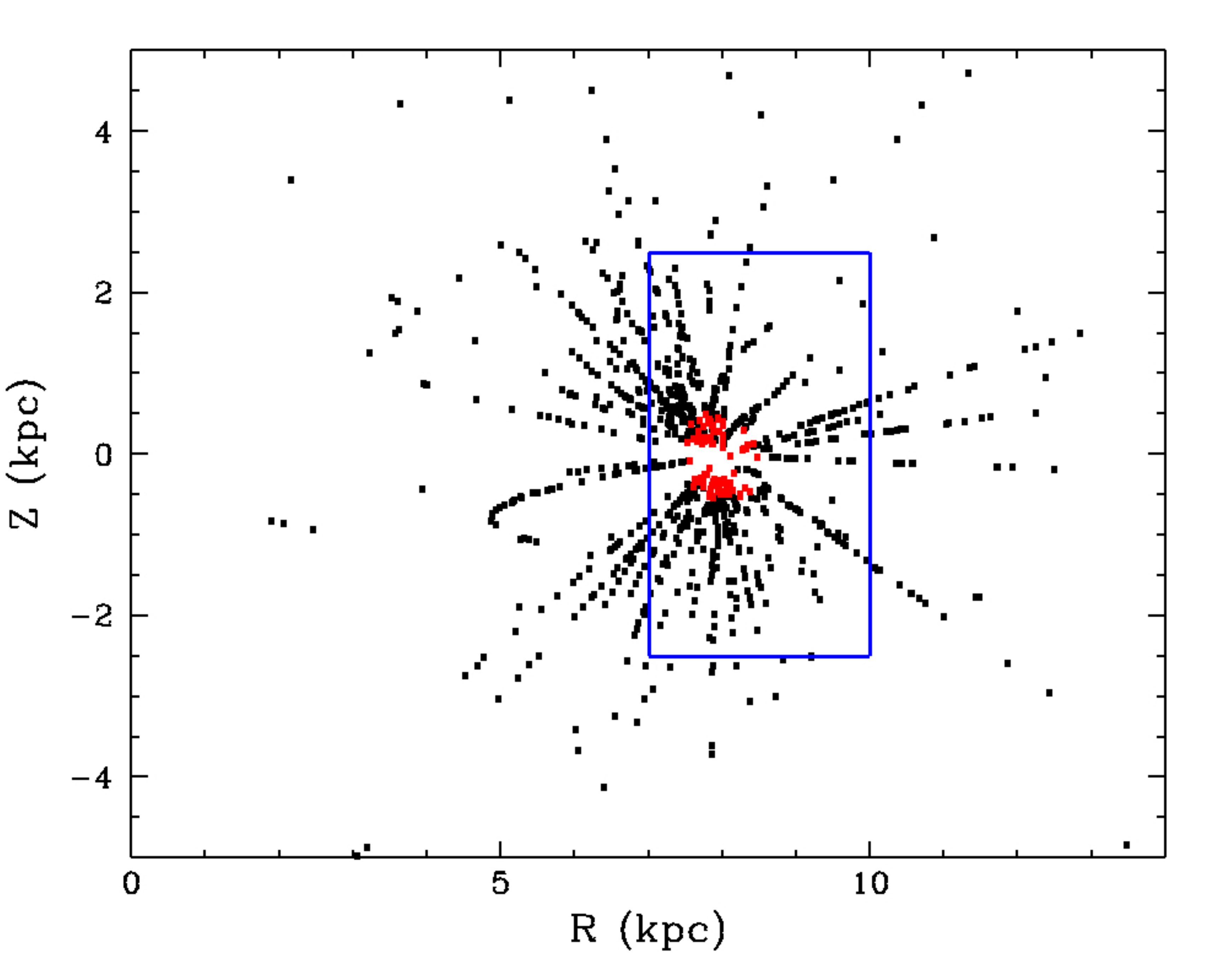}  
\caption{Target's location, for sub-sample {\it e} of Fig.~\ref{Pops}, in 
vertical distance to the Galactic plane (Z) and radial cylindrical Galactic coordinate (R). Red points indicate the stars
with a distance value smaller than 600~pc from the Sun (10$\%$ of the sample). The blue rectangle shows the maximum R and Z 
values of the \citet{Lee2011b} analysis of SEGUE data.}
\label{RZ}
\end{figure}

First of all, we have compared the spatial distribution of the analysed targets
with that of previous spectroscopic studies of large numbers of disc stars, with available measurements of the \aabun \ ratio,
as it is the case of the \citet{Lee2011b} and \citet{Adibekyan2012} analysis. We recall that the former is a low-resolution 
study of stars with Galactic radius in the range 7$-$10~kpc, and the latter is a high-resolution analysis of solar-neighbourhood stars,
located at a maximum distance of about 600~pc from the Sun. Fig.~\ref{RZ} shows the location
of the GES targets analysed in this paper ({\it e} sub-sample of Fig.~\ref{Pops}), in vertical distance to the Galactic plane and
radial cylindrical Galactic coordinate {\it R}. To understand how many of the targets are outside the solar neighbourhood, 
the points are colour coded by distance. All the objects with distances smaller than D$=$600~pc from the Sun are plotted in red and those with larger distances in black. As expected from the photometric target selection, 90\% of the stars are at 
distances larger than D$=$600~pc from the Sun (i.e. much further than the  \citet{Adibekyan2012} sample, comprised mainly of stars
within 50~pc).

Regarding the thin to thick disc separation in the \aabun~ vs. \met~ plane, the above identified gap 
is in agreement with the separation seen in the \citet{Adibekyan2012} data.
Fig.~\ref{Pops} shows the \citet{Adibekyan2012} division, as a red line with three segments, that
can be compared to our proposed separation, in black.   Both lines are compatible, with very
small differences probably due to possible offsets in the abundance measurements between the GES analysis and the 
\citet{Adibekyan2012} one, or to a possible adjustment between the [M/H] and the [Fe/H] abundances. On the other hand, 
the \citet{Lee2011b} analysis proposes a division line with a shallower
slope as illustrated by the blue solid line in Fig.~\ref{Pops}. The two dashed lines indicate their dividing points for the 
thin and the thick disc stars to avoid the misclassification of stars. We see that the lower dashed line allows to select
thin disc stars in fair agreement with both our selection and the \citet{Adibekyan2012} one. However, the upper dashed line
excludes an important part of the thick disc sequence with intermediate and low \aabun~ values, even if a bias between our \aabun~
measurements and the \citet{Lee2011b} ones could exist. This could come mainly from the identification of 
the metal-rich part of thick disc, that would be more difficult to separate from the thin disc from low resolution data, 
due to the small differences in the \aabun \ values.

Furthermore, the difference of slope observed in our sample between the thin and the thick disc sequences seems in agreement
with the \citet{Haywood2013} analysis of \citet{Adibekyan2012} data, that already points out an iron enrichment of the thin
disc lower than the thick disc one.

Finally, concerning the metal-rich end of the thin disc, no $\alpha$-rich metal-rich stars are observed in our sample, 
nor in the \citet{Lee2011b} one, contrary to the results of \citet{Adibekyan2011} and \citet{Gazzano13} and \citet{Boeche}, 
restricted to, or predominantly in, smaller distances from the Sun. Furthermore, we note here that the number of stars with 
\met \ higher than -0.1~dex is rather small in our sample, compared to the previously mentioned nearby studies.

\section{Distribution of distances to the Galactic plane}

\begin{figure*}[th!]
\centering
\begin{tabular}{c c}
\includegraphics[width=8.5cm,height=8cm]{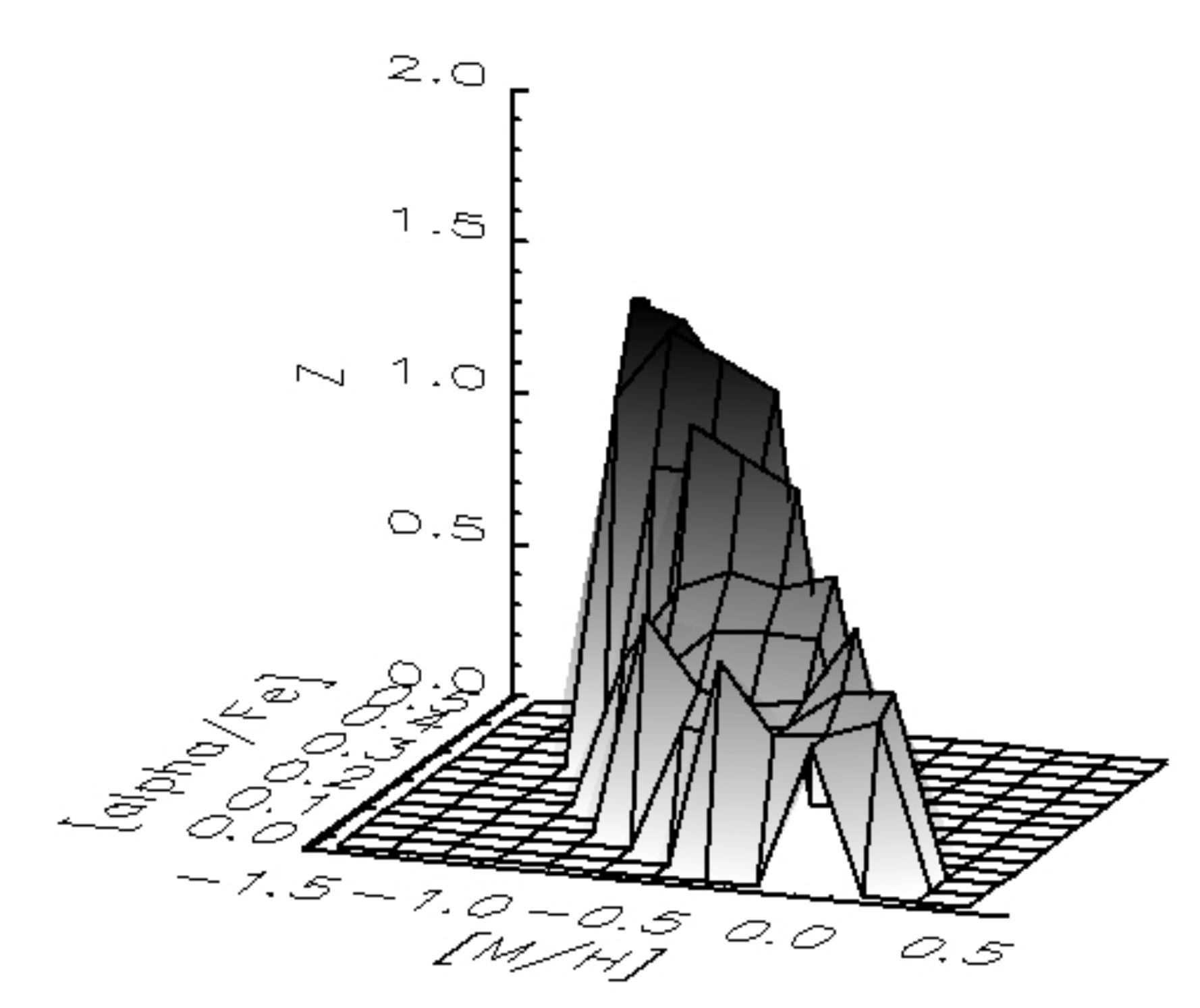}  & \includegraphics[width=8.5cm,height=8cm]{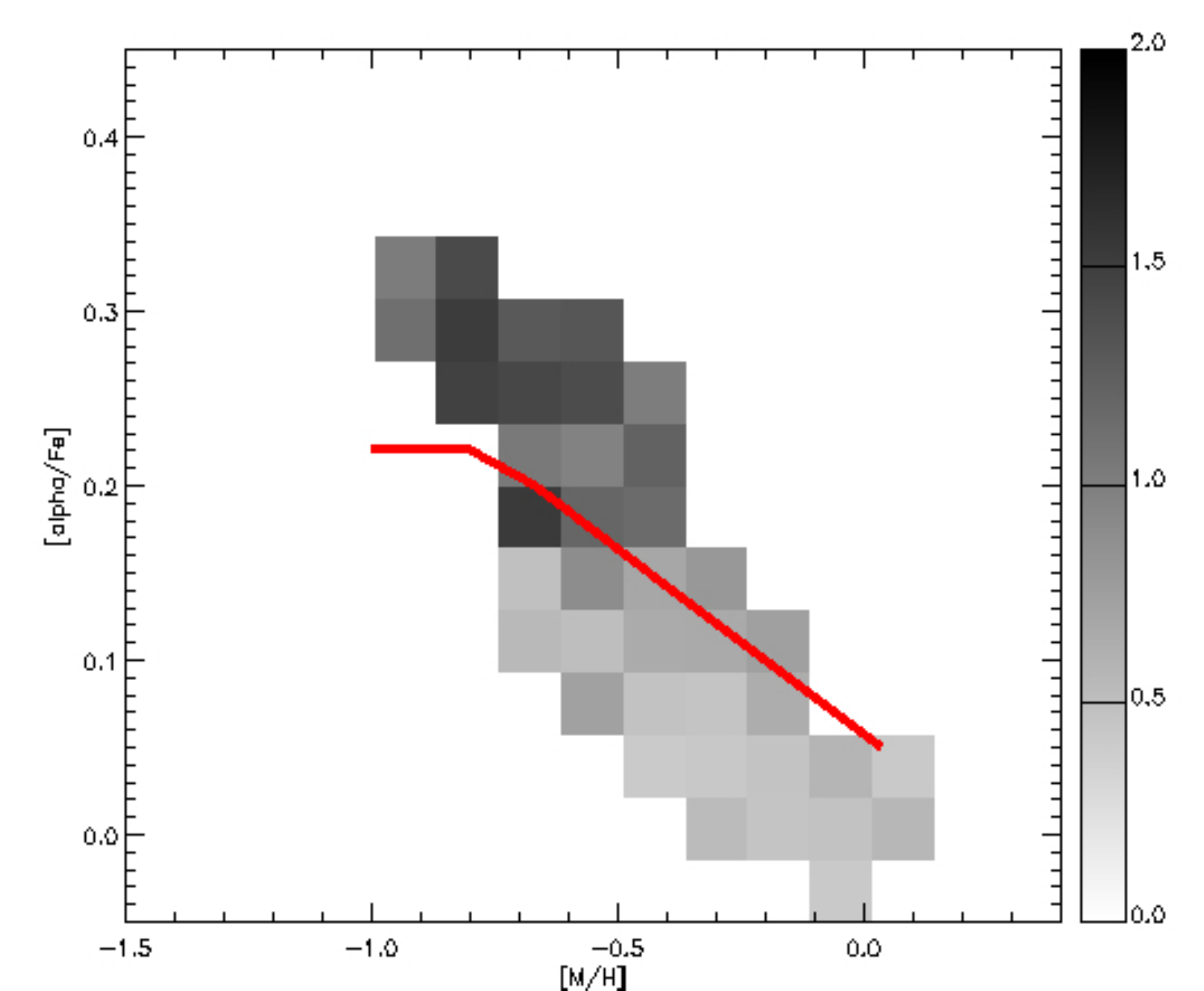} 
\end{tabular}
\caption{Left panel: three-dimensional plot of the distance to the Galactic plane, in the z-axis, shown as a function of \met~ and \aabun~ for the analysed stars (sub-sample $e$ of Fig.~\ref{Pops}). Right panel: Distribution of the distances to the Galactic plane in \aabun \ vs. \met. The red line shows the separation between the thick and the thin discs sequences defined in Sect. 4.
In this panel, high-Z values are coded in darker grey than lower Z values.}
\label{Zsurfaces}
\end{figure*}

As already mentioned in the Introduction, the thin and the thick discs are characterised by different
scale heights. In order to explore the transition between these two populations, keeping in mind their chemical
separation in the \aabun~ vs. \met~ plane as described in the above section, we have analysed the distribution of the derived 
distances to the Galactic plane (from the absolute values of the Galactic Z coordinate) for the target stars. 

The left panel of Fig.~\ref{Zsurfaces} shows a three-dimensional view of the distance to the Galactic plane, 
in the Z-axis, as a function of \met\ and \aabun, for the analysed stars (sub-sample {\it e} of Fig.~\ref{Pops}).
This illustration is complemented, in the right panel, with a grey scale-coded two-dimensional plot of the distribution of 
{\it Z} values in the \aabun~ vs. \met~ plane. The red line shows the separation between the thick and the thin
discs sequences, defined in Sec. 4 (c.f. Fig.~\ref{Pops}). We point out that the possible halo stars discussed 
in the previous section are not shown in Fig.~\ref{Zsurfaces}, because of the low number of objects per bin.

Regarding the thick disc population, the mean distances to the Galactic plane seem to decrease along the sequence, from the metal-poor
end to the metal-rich one: 
\begin{itemize}
\item Z$\sim$1.5~kpc at \met$\sim$-0.9~dex and 0.28$<$\aabun$<$0.40~dex
\item Z$\sim$1.2~kpc at \met$\sim$-0.5~dex and 0.21$<$\aabun$<$0.30~dex
\item Z$\sim$0.8~kpc at \met$\sim$-0.3~dex and 0.12$<$\aabun$<$0.25~dex. 
\end{itemize}

Therefore, as the metallicity increases, the thick disc seems to lay in thinner and thinner layers.
On the other hand, the distribution of vertical distances to the plane with respect to the
Galactocentric radial distance is not perfectly homogeneous (c.f. Fig.~\ref{RZ}). More particularly,
our sample is dominated by stars further away from the galactic plane in the inner parts of the disc with
respect to the typical heights probed in regions outside of the solar radius.
Thus,   different   Z   heights   can be   representative   of   different   ranges   in   R, and,
if   there   were   significant   metallicity   variations   with   respect   to   R,   this   would   
manifest   as  a  variation   in   metallicity   with   height.  However, as later explained in
Sec.~9.2, our data show a flat distribution of \met \ and \aabun \ with galactocentric radius for the thick disc. 
Therefore, the gradient of Z that appears along the thick disc sequence in the \met \ v.s. \aabun \ plane,
for increasing \met \ values should be   a   real   effect,   not   caused   by   sampling.

On the other hand, the thin disc sequence shows a quite constant value of the mean distances of its stars
to the Galactic plane of {\it Z}$\sim$0.5~kpc, from the metal-poor side (\met~ $\sim$-0.8~dex) to the metal-rich one 
(\met~ $\sim$0.3~dex). Again, this conclusion should not be influenced by the spatial coverage of the studied
sample, despite the fact that the thin disc has a metallicity gradient with respect to R (c.f. Sec.~9.2). This is
because the thin disc Z range is fairly homogeneously covered at the full range of Galactocentric radius, and, in
particular, the mean Z value of the stars in the thin disc sequence does not present any clear trend with R.

As a consequence of the above conclusions, the transition from the thick disc sequence to the thin disc one,
in the \aabun \ vs. \met plane, implies a steeper change in the mean Z distances at [M/H] around -0.8~dex than 
around -0.3~dex. This coincides with the fact that the chemical separation between both sequences (by a gap or a 
lower density region) seems clearer in the metal poor regime than in the metal rich one.

Finally, we point out that, as expected, there is a correlation between the Z distances and the associated errors, with larger errors for stars that are further from the plane. More particularly, the stars in the metal-poor high-alpha end ([M/H] around -1.0~dex) have errors around 0.45 kpc, while at [M/H] around -0.7~dex the error is about 0.20~kpc for the thick disc stars and around 0.075 kpc for the thin disc. Nevertheless, while the stars that are further from the plane have larger absolute distance errors, the percent
error is actually smaller, as indicated in Figures~\ref{errordistMet} and \ref{errordistDist}.



\section{Distribution of rotational velocities}

\begin{figure}[th!]
\includegraphics[width=8.5cm,height=8cm]{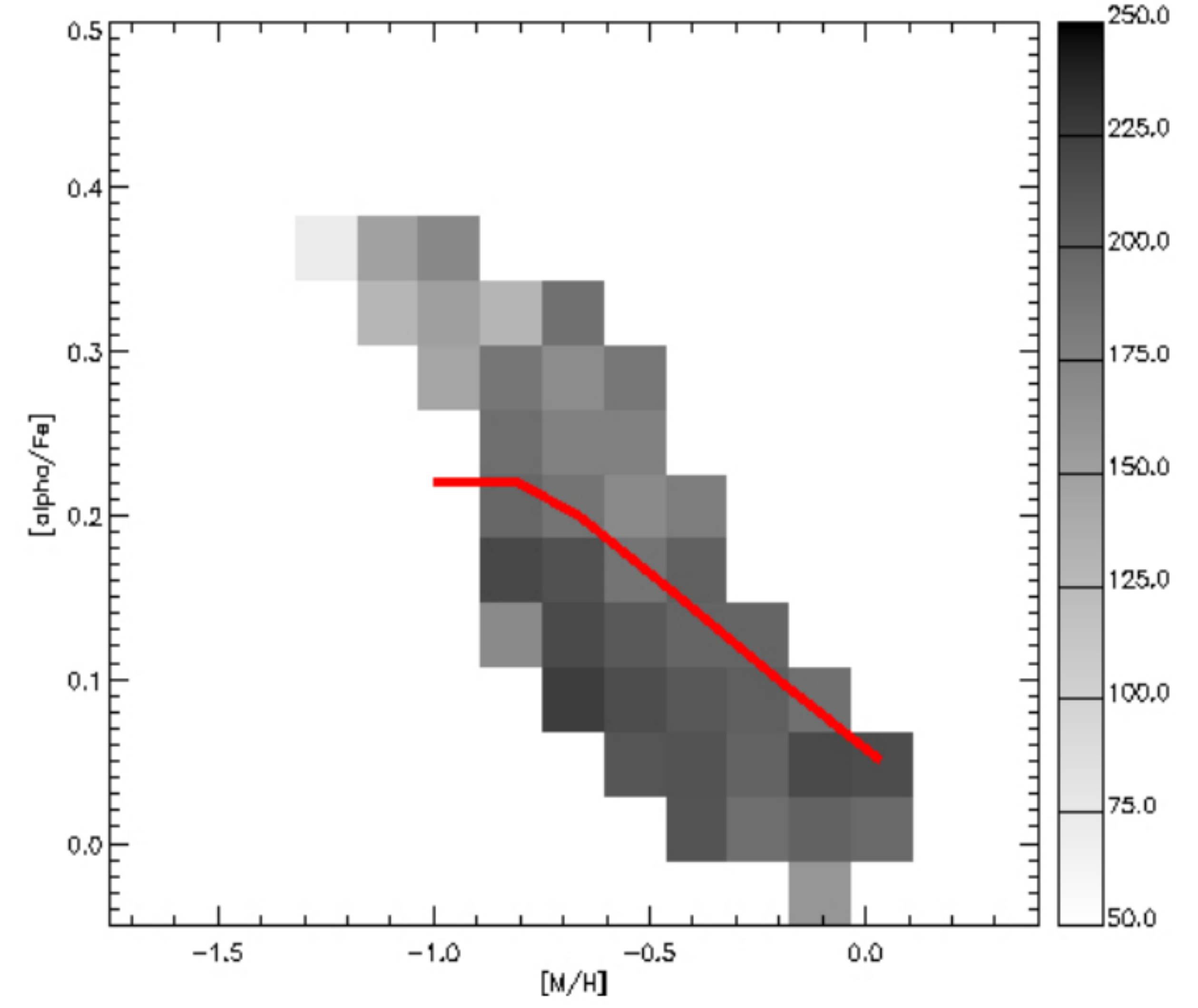}  
\caption{Distribution of the cylindrical rotational velocity, V$\Phi$, in the \aabun~ vs. \met~ plane. Large velocities 
values are coded in darker grey than lower ones. As in Fig.~\ref{Zsurfaces}, the red line shows the separation between the thick and the thin discs.}
\label{Vphi}
\end{figure}

\begin{figure}[th!]
\includegraphics[width=9cm,height=7cm]{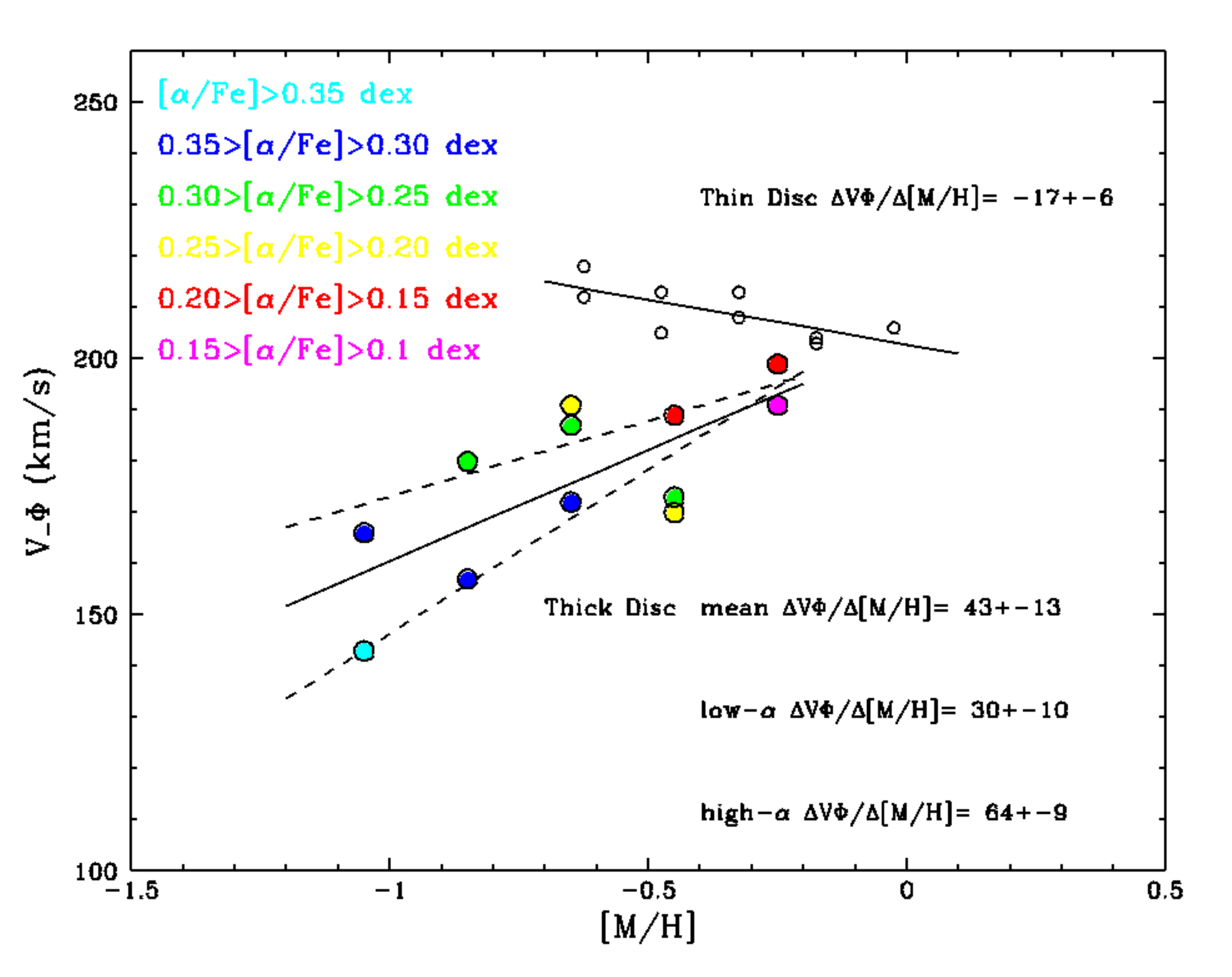}  
\caption{Rotational velocity gradients with metallicity for the thin (open circles) and thick (filled circles) sequences.
The thick disc points are colour coded by \aabun \ abundance intervals.
The Spearman's rank correlation coefficient between V$_{\Phi}$ and \met \ is 0.73.}
\label{RotMetGrad}
\end{figure}

The distribution of the cylindrical rotational velocities, V$_{\Phi}$, in the \aabun~ vs. \met~ plane, is
shown in Fig.~\ref{Vphi}.
 The rotational velocities of the stars located in the thick disc sequence have a mean
value of V$_{\Phi}=$176$\pm$16~km/s, and those in the thin disc have  a mean V$_{\Phi}=$208$\pm$6~km/s. In addition, for the thick disc,
the V$_{\Phi}$ values seem to progressively increase from the metal-poor
$\alpha$-rich end (with V$_{\Phi}\sim$150~km/s) to the metal-rich $\alpha$-poor one (V$_{\Phi}\sim$190~km/s), reaching values close to the thin
disc ones.

On the other hand, a correlation between rotational velocity and metallicity is observed for both the thin and the thick
disc stars, although with opposite signs. In order to exploit the fact that both \aabun \ and \met \ are available for
our sample, we have decided to analyse the correlation of rotational velocity taking into account both chemical abundance parameters.
To this purpose, and as illustrated in Fig.~\ref{RotMetGrad}, we have used the mean values of the rotational
velocity, for different bins of 0.05~dex in \aabun \  and 0.20~dex in \met, to derive its dependence with metallicity.
The chosen bin lengths are higher than the mean errors in \aabun~ and \met~ (0.03~dex and 0.08~dex, respectively),
but they allow us to have a robust statistics per bin (a mean of 64 stars for the thin disc sequence and 34 stars 
for the thick disc one).  

First, all the bins along the thin and thick disc sequences were considered.
For the thin disc stars, we find a gradient of Galactic rotation with metallicity of -17$\pm$6~km~$s^{-1}$~dex$^{-1}$.
We have checked the possible influence of the large range of sounded galactocentric radius on the derived gradient, by
splitting the sample in two: stars inwards (R$<8kpc$) and outwards (R$>8kpc$) of the solar radius. The resulting values
of the gradient are -20$\pm$13 and -22$\pm$9~km~$s^{-1}$~dex$^{-1}$, suggesting that the derived slopes are not indirectly
created by radial gradients in metallicity.
For the thick disc population, we find a positive gradient of 43$\pm$13~km~s$^{-1}$~dex$^{-1}$. The derived values of the
gradient for the thin and the thick disc are in very
good agreement with the \citet{Lee2011b} analysis of SEGUE data (who find -22.6$\pm$1.6~km~s$^{-1}$~dex$^{-1}$
for the thin disc and 45.8$\pm$1.8~km~s$^{-1}$~dex$^{-1}$ for the thick disc) and with the \citet{Adibekyan2013}
measurements in the solar neighbourhood (-16.8$\pm$3.7~km~s$^{-1}$~dex$^{-1}$
for the thin disc and 41.9$\pm$18.1~km~s$^{-1}$~dex$^{-1}$ for the thick disc). Our gradient for the thick disc
is also in agreement with the measurements of \citet[][45$\pm$12~km~s$^{-1}$~dex$^{-1}$]{Kordopatis11b}  and \citet{Spagna2010},
derived, respectively, from low resolution FLAMES spectroscopy and SDSS spectro-photometry, with no
estimations of \aabun. It also confirms the disagreement with the SDSS result
of \citet{Ivezic}, based on photometric metallicities.

Second, to analyse if the $\Delta$V$_{\Phi}/\Delta$\met \ value depends on the considered bins of \aabun, 
we have derived the mean value of \aabun~ for each metallicity bin and divided the thick disc sample in two regimes:
\begin{itemize}
\item a high-$\alpha$ sample with those bins corresponding to \aabun~ values higher than the mean \aabun \
for a given metallicity
\item a low-$\alpha$ sample with the bins corresponding to lower \aabun~ values than the mean.
\end{itemize}

The mean values of \aabun \ along the thick disc sequence dividing the above defined 
sub-samples, at each metallicity bin, are 0.35~dex at \met$=-$1.0~dex, 0.30~dex at \met$=-$0.85~dex, 0.25~dex at \met$=-$0.65~dex,
0.20~dex at \met$=-$0.45~dex and 0.15~dex at \met$=-$0.25~dex. 
The corresponding derived $\Delta$V$_{\Phi}/\Delta$\met \ slopes are overplotted in Fig.~\ref{RotMetGrad}. The gradient of
rotation with metallicity seems to be steeper than the mean (64$\pm$9~km~s$^{-1}$~dex$^{-1}$ instead of
43$\pm$13~km~s$^{-1}$~dex$^{-1}$) when the high-$\alpha$ bins are considered. On the contrary,
the value of $\Delta$V$_{\Phi}/\Delta$\met \ is lower than the mean one when it is determined from
the low-$\alpha$ bins (30$\pm$10~km~s$^{-1}$~dex$^{-1}$). 
Given that the \aabun~ dispersion is rather comparable to the measurement errors, 
the existence of a real intrinsic dispersion can not be concluded. Indeed, the described dependence of the
$\Delta$V$_{\Phi}/\Delta$\met \ gradient with the considered \aabun~ bins could be the sign of a
contamination of the thick disc stars from the thin disc ones in the lower part of the thick
disc sequence. If this is correct, the steeper gradients derived from the high-$\alpha$ bins
could be more characteristics of the thick disc population than the previously published ones
of about 43$\pm$13~km~s$^{-1}$~dex$^{-1}$. To test this possibility, we have also derived the value of the slope,
using all the bins along the thick disc sequence with a mean error in \aabun \ lower than 0.04~dex. The resulting slope
is 56$\pm$19~km~s$^{-1}$~dex$^{-1}$ instead of 43$\pm$13~km~s$^{-1}$~dex$^{-1}$, confirming that the slope seems steeper 
for higher quality data.

\begin{figure}[th!]
\includegraphics[width=9cm,height=7cm]{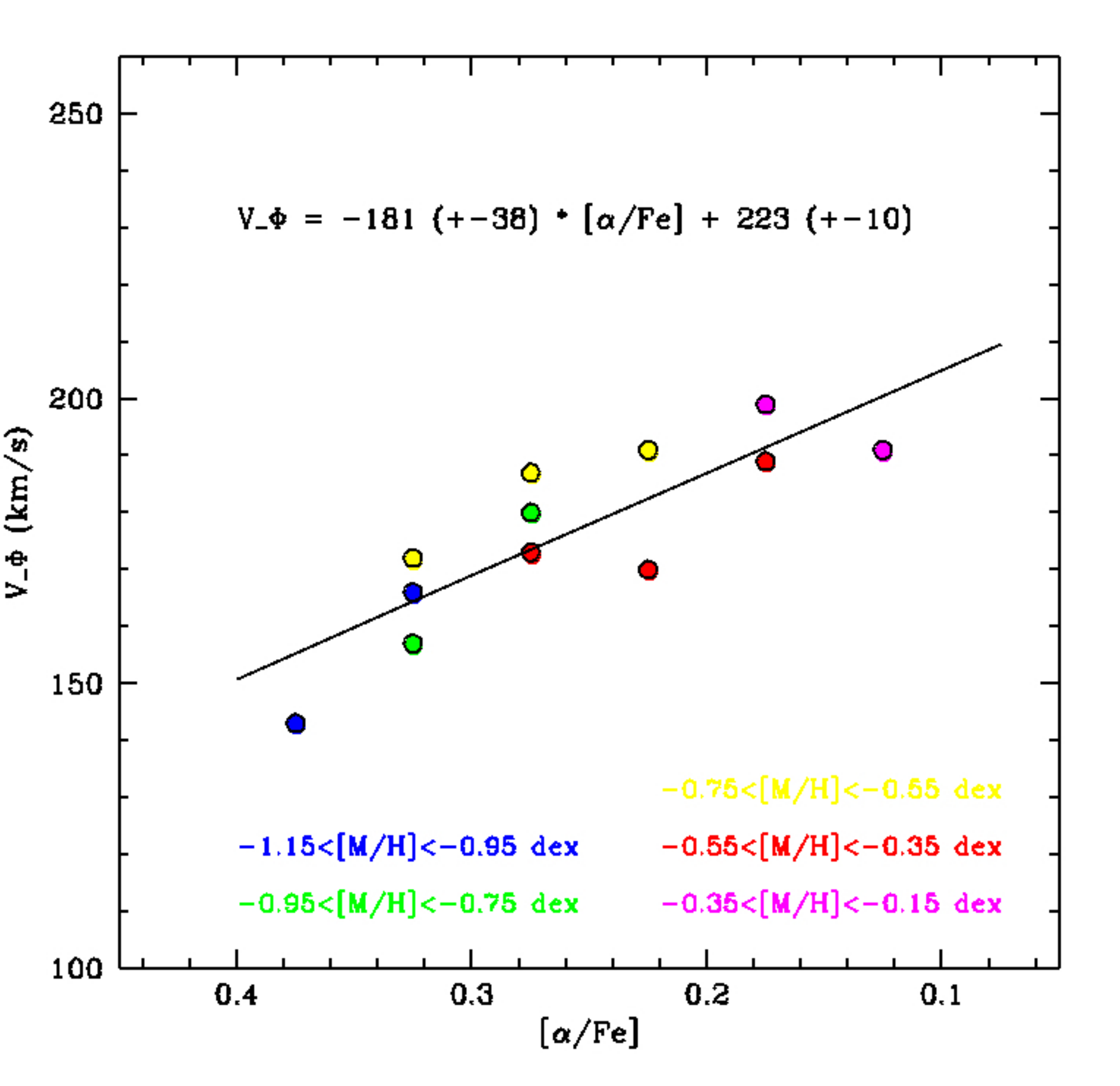}  
\caption{V$_{\Phi}$ as a function of \aabun for the thick disc sequence stars. The points correspond to the same bins in
\aabun \ and \met \ than those of Fig.~\ref{RotMetGrad} and they are colour coded by metallicity. The Spearman's rank 
correlation coefficient between V$_{\Phi}$ and \aabun~ is -0.85.}
\label{RotAlf}
\end{figure}

Finally, to analyse the direct influence of \aabun \ in the stellar rotation for thick disc stars, we have also derived
the correlation between rotation and \aabun \, separate from \met. As shown in Fig.~\ref{RotAlf}, V$_{\Phi}$ and \aabun~
are highly anticorrelated (the Spearman's coefficient value is -0.85, corresponding to 0.72\% of the V$_{\Phi}$ variance). 
In addition, the correlation rank between V$_{\Phi}$ and \aabun~ is even more significant than the one between V$_{\Phi}$ and \met \ 
(having a Spearman's coefficient value of 0.73, that accounts for 0.53\% of the V$_{\Phi}$ variance).

\begin{figure}[th!]
\includegraphics[width=9cm,height=7cm]{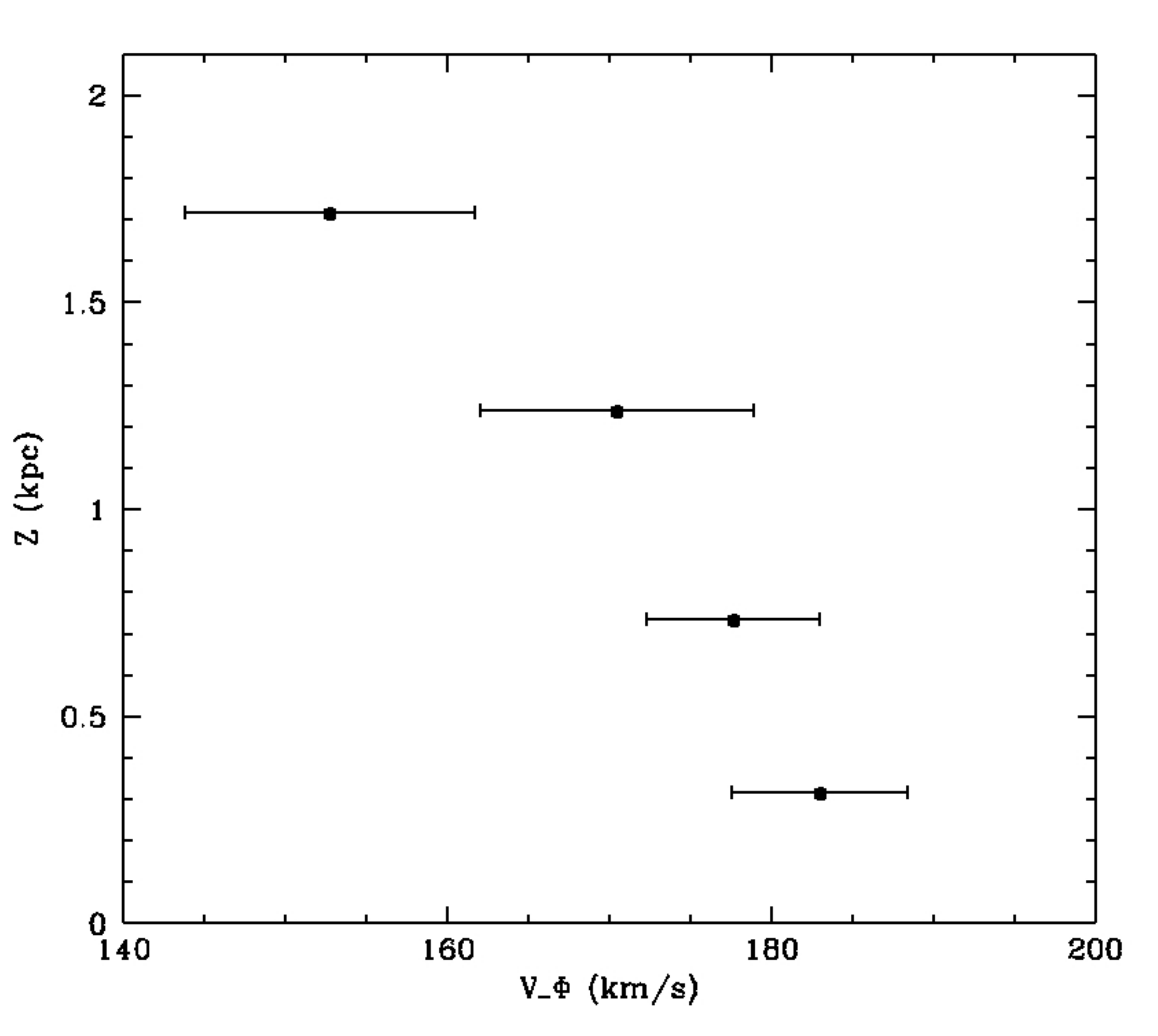}  
\caption{Mean V$_{\Phi}$ value as a function of vertical height for the thick disc sequence stars.}
\label{RotZ}
\end{figure}

Furthermore, we have checked how azimuthal velocity varies with vertical height for the thick disc sequence stars. 
Fig.~\ref{RotZ} shows the mean V$_{\Phi}$ value as a function of Z, in four bins of 0.5~kpc. The error bars
indicate the standard error of the mean, in each bin. Clearly, the azimuthal
velocity seems to decrease with height, in particular for Z values higher than about 1~kpc. As a consequence,
this could explain why in situ samples of the thick disc can show a higher value of the thick disc lag (hence lower V$_{\Phi}$
values) than the local surveys \citep[see for instance][]{WyseLag}. 

\section{Distribution of azimuthal and vertical velocity dispersions}

The stellar velocity dispersions can also be studied for the two chemically separated thin and thick
disc sequences with our data. To this purpose,  as low quality 
velocity determinations increase artificially the velocity dispersions,
we have further restricted our analysis sub-sample, allowing
only stars with errors in azimuthal and vertical velocity lower than 70 km/s. 
This value was found imposing a good balance between the statistical robustness and the 
allowed maximum error in velocity. This new restriction, in addition 
to those of sub-sample {\it e} presented in Table \ref{Samples}, leaves a data set of 645 stars, 
with a mean error in V$_{\Phi}$ of 31~km/s.

\begin{figure*}[th!]
\centering
\begin{tabular}{c c}
\includegraphics[width=8.5cm,height=8cm]{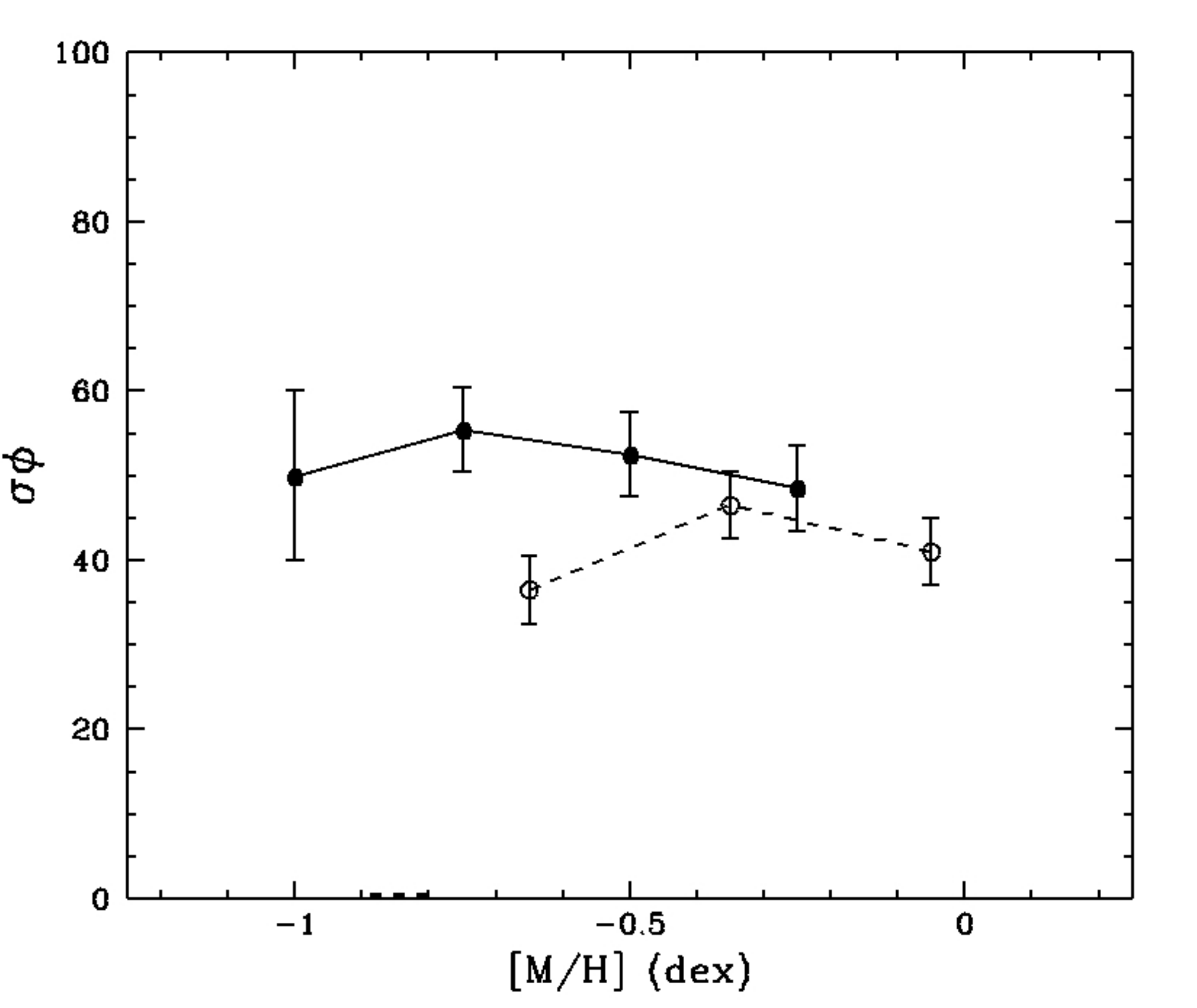}  & \includegraphics[width=8.5cm,height=8cm]{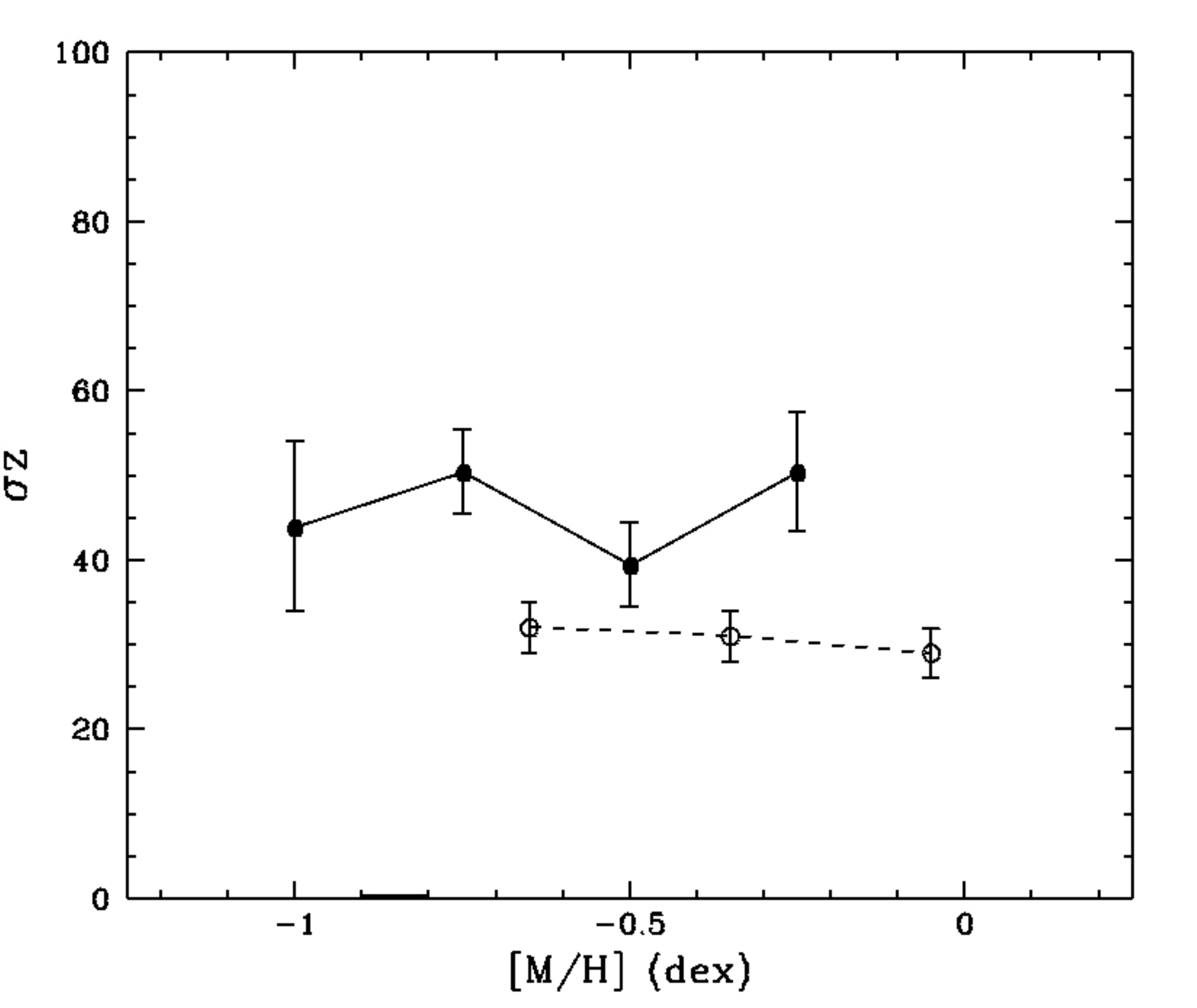}\\
\includegraphics[width=8.5cm,height=8cm]{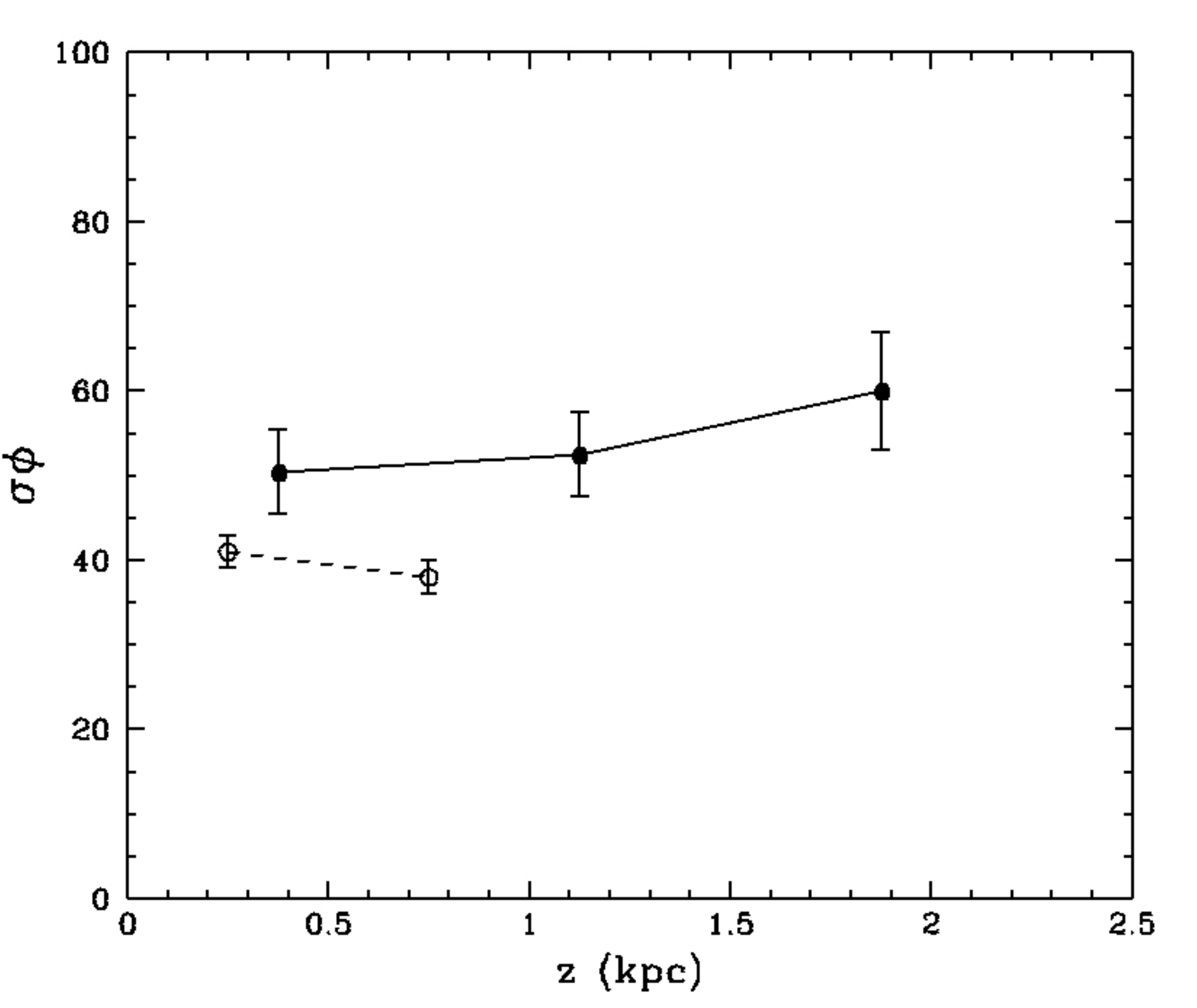}  & \includegraphics[width=8.5cm,height=8cm]{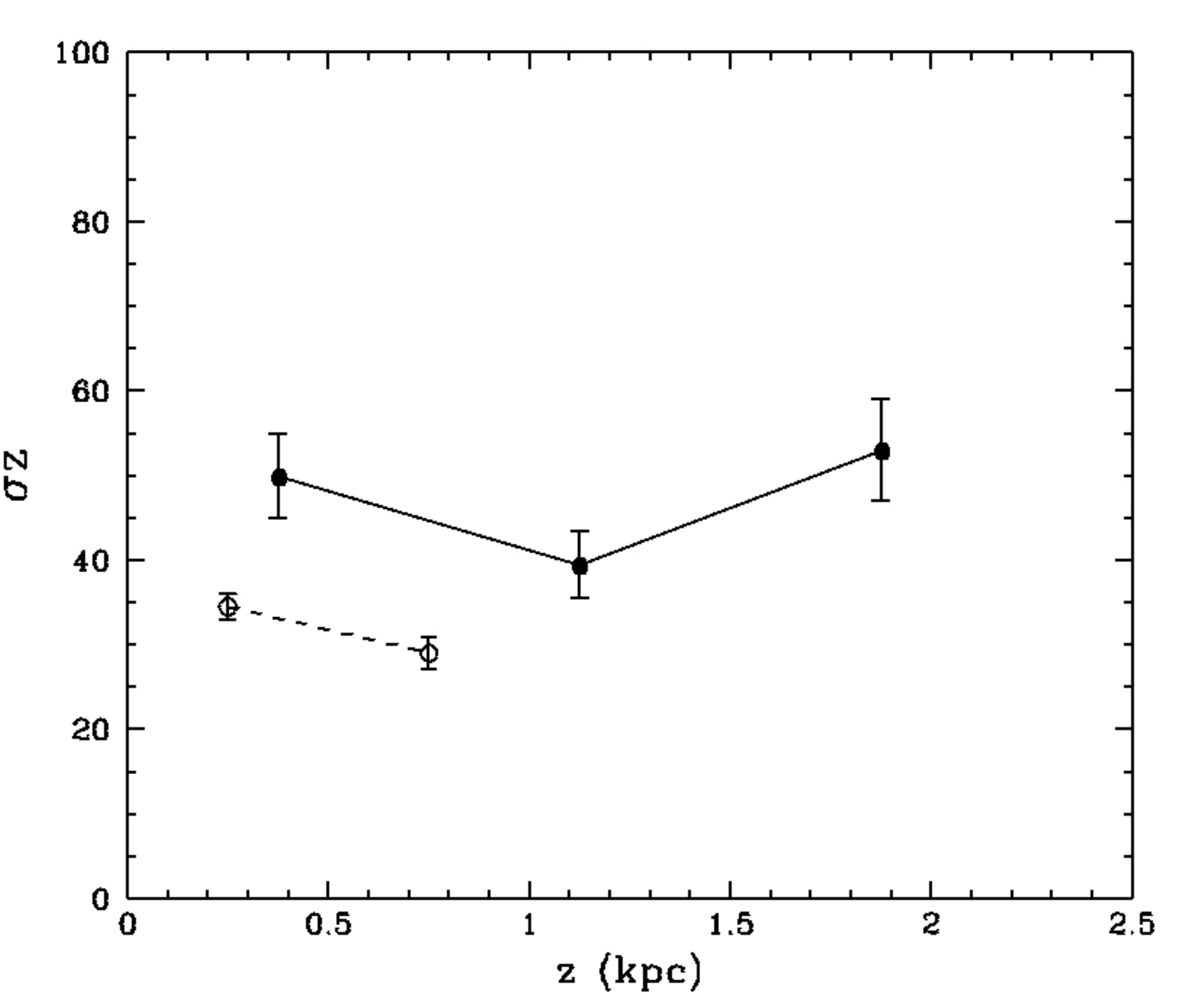} 
\end{tabular}
\caption{$\sigma_{\Phi}$ (left panels) and $\sigma_{Z}$ (right panels) as a function of metallicity (upper panels) and 
distance to the Galactic plane (lower panels). Filled circles correspond to the chemically selected thick disc sequence and 
open circles to the thin disc sequence.}
\label{Sigmas}
\end{figure*}

 Fig. \ref{Sigmas} shows the distribution of the measured (not error free) azimuthal and 
vertical velocity dispersions ($\sigma_{\Phi}$ and $\sigma_{Z}$) as a function of metallicity and distance
to the Galactic plane. The error bars come from the standard error of the derived standard deviations. 
As expected, both velocity dispersions are higher for the thick disc sequence than for the thin
disc one. The mean value of $\sigma_{\Phi}$ is 54$\pm$5~km/s for the thick disc and 40$\pm$3~km/s for the thin disc. Regarding
$\sigma_{Z}$, the mean values are 48$\pm$5~km/s for the thick disc sequence and 32$\pm$3~km/s for the thin disc one.
As shown in the upper panels of Fig. \ref{Sigmas}, no significant dependencies of either $\sigma_{\Phi}$ or $\sigma_{Z}$ with
metallicity or Z seem to be present in the data. Nevertheless, for the thick disc, the measurements suggest a possible increase 
of $\sigma_{\Phi}$ with distance to the Galactic plane that is not visible for $\sigma_{Z}$. This conclusion seems strengthened
by the increase of the uncertainty on the derived mean V$_{\Phi}$ values as Z increases, as shown in Fig.~\ref{RotZ}. 
A possible decrease of $\sigma_{\Phi}$
with increasing metallicity could also be present for the thick disc. In addition, around \met$\sim$-0.3~dex, the dispersion
in the azimuthal velocity of the thick and the thin discs seem roughly the same, with a further decrease of $\sigma_{\Phi}$
for the metal rich end of the thin disc. This seems to be in agreement with the behaviour of the azimuthal velocity as a function of the
metallicity, shown in Fig. \ref{RotMetGrad} (see also Section~6). Moreover, the metal-poor side of the thin disc sequence, that
presents higher rotation values, shows also a slightly lower azimuthal velocity dispersion.

On the other hand, we point out that our determinations of the velocity dispersions could be affected by the lack of stars at very low
Z values (only 37$\%$ of the thin disc stars and 17$\%$ of the thick disc ones, included the first considered bin of vertical distance, 
are at abs(z)$<$250~pc). This could explain the slightly higher values of dispersion found close to the Galactic plane with respect to
other surveys like RAVE \citep[e.g.][]{Burnett}.

Finally, for the thick disc sequence, the azimuthal velocity dispersion seems to be of the order of the vertical velocity dispersion, within
the error bars. In any case, the analysed data do not suggest, as predicted by \citet{Binney}, a smaller $\sigma_{\Phi}$ than the
corresponding $\sigma_{Z}$ for the thick disc.


\section{Orbital parameters}

To calculate the orbital parameters of the stars in the sample we integrated orbits in a Galactic
potential consisting of a \citet{Miyamoto} disc, a \citet{Hernquist} bulge and a spherical logarithmic
halo. Their characteristic parameters took the values $M_{disk} = 9.3 \times 10^{10} M_\odot$ and 
 $a = 6.5$ kpc, $b = 0.26$ kpc for the disk, $M_{bulge} = 3 \times 10^{10} M_\odot $ and $c = 0.7$ kpc for the bulge, and 
$v_h = 134.5$ km/s, and $d = 12$ kpc for the halo. In particular, the eccentricities of the stars were 
defined as $(r_{apo} - r_{peri}) / (r_{apo} + r_{peri})$ where $r_{apo}$ ($ r_{peri}$)  is the maximum
(minimum) distance reached by the star in its orbit in the past 2~Gyr.

\subsection{Eccentricity distribution}

\begin{figure}[th!]
\includegraphics[width=8.5cm,height=8cm]{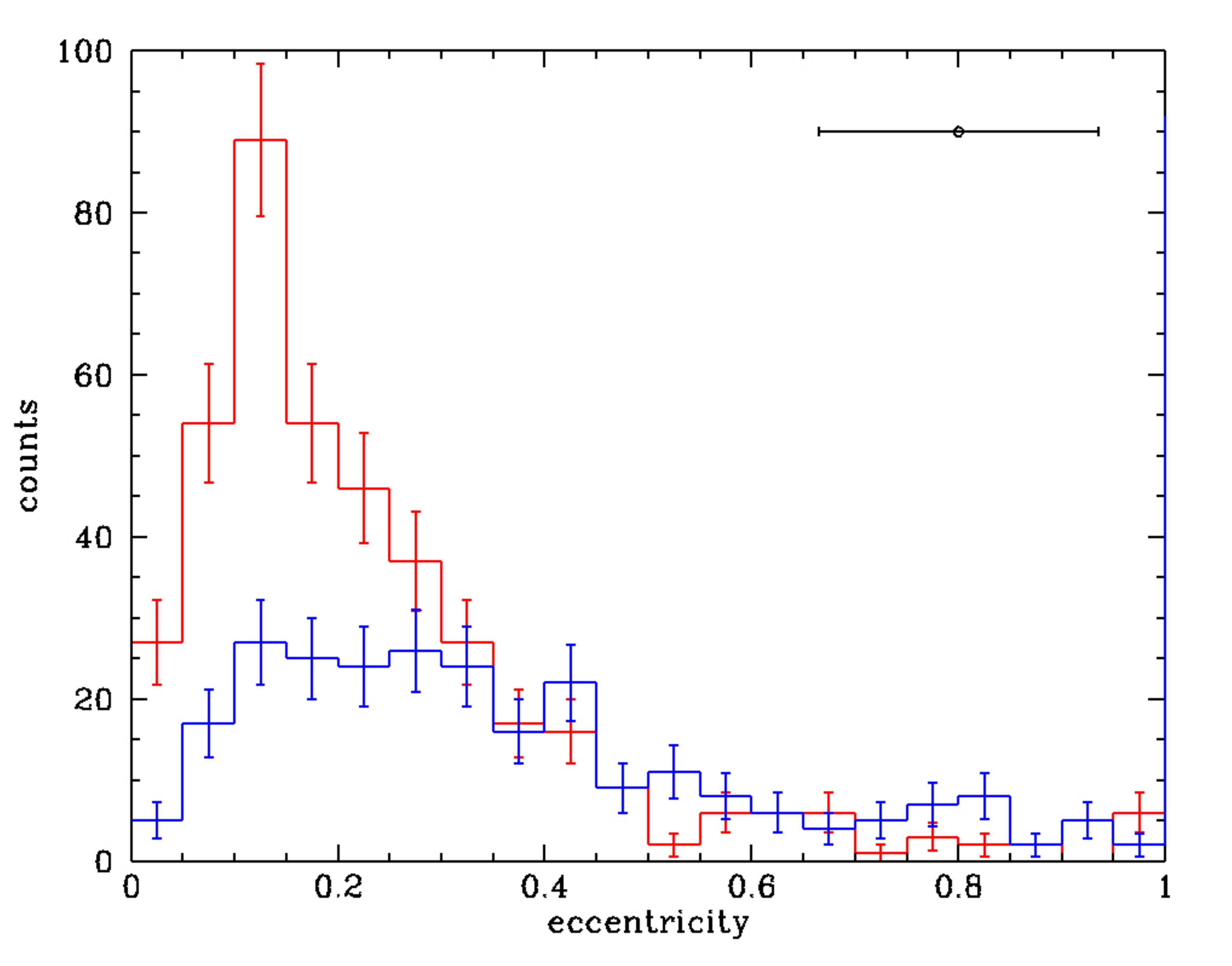}   
\caption{Eccentricity distributions, with the Poisson uncertainties, for the thin (red) and thick disc (blue) sequences. The typical error bar for the eccentricities is shown on the top right corner of the figure.}
\label{Eccentricity}
\end{figure}

\begin{figure}[th!]
\includegraphics[width=8.5cm,height=8cm]{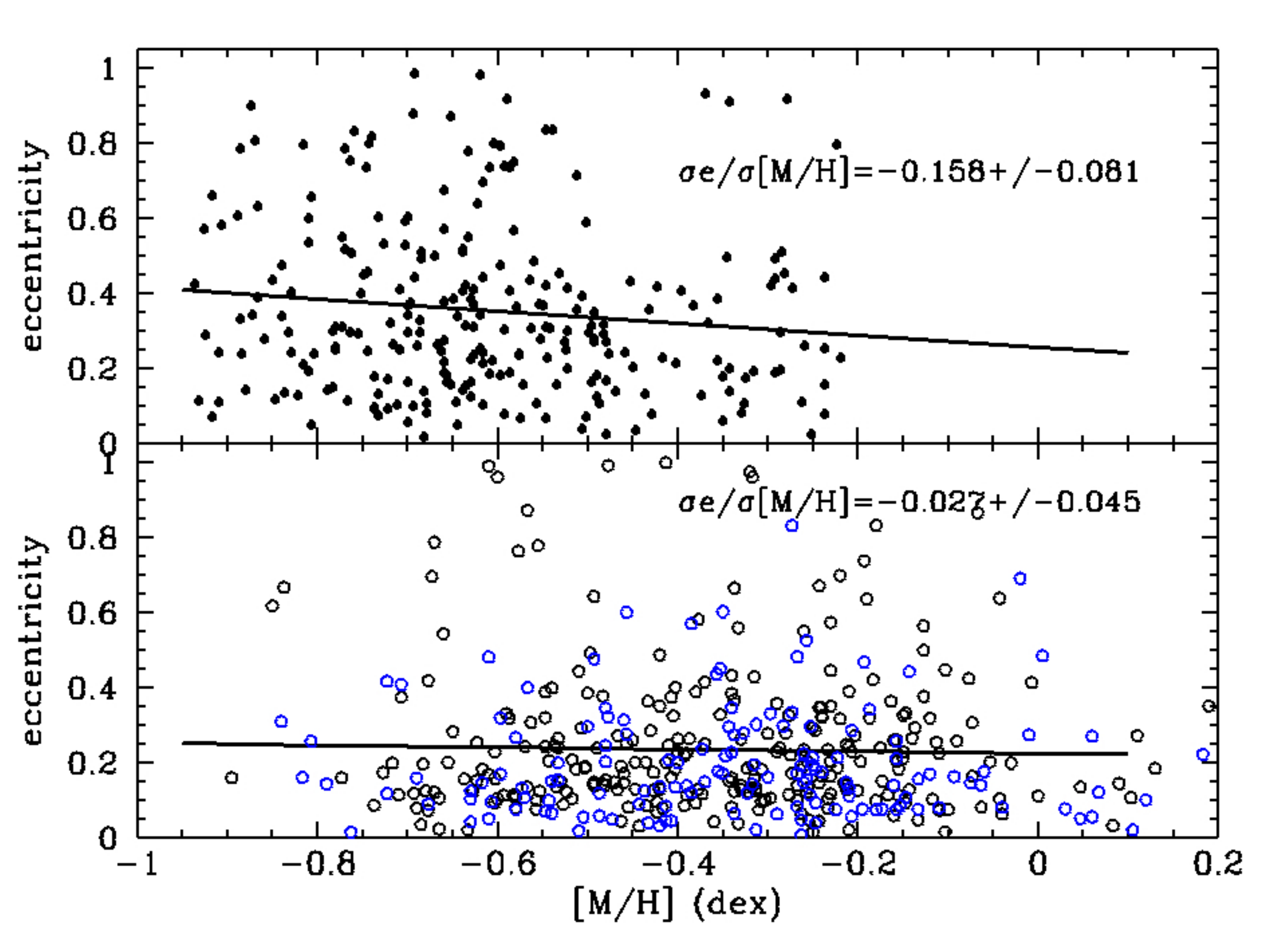}   
\caption{Stellar orbital eccentricities as a function of metallicity for the thick (upper panel) and the thin (lower panel) discs. In the lower panel, the thin disc stars with distances to the Galactic plane lower than 400~pc are plotted as blue open circles.}
\label{EccentricityMet}
\end{figure}

It is today widely accepted that the shape of the eccentricity distribution of the stellar
orbits can provide important constraints on the formation mechanisms of the Galactic disc \citep{Sales09}.
Figure~\ref{Eccentricity} shows the distribution of orbital eccentricities for the 
thin (red line) and the thick disc (blue line) chemically selected stars. 
 As expected, the thin disc distribution is strongly peaked at a low eccentricity value of $\sim 0.15$. On the other hand, the thick disc
is significantly broader and its median eccentricity is around 0.3. Compared to previous determinations, such as those 
of \citet{Wilson11}, \citet{Lee2011b} and \citet{Kordopatis11b}, the shape is flatter and does not decline 
as fast at high eccentricities, although because of the small number of stars, the significance of this may be questionable.

In order to evaluate the errors in the computed eccentricities, we have 
calculated the median, 25$\%$ and 75$\%$ percentile values of the eccentricity from all orbits starting 
from 100 realisations of the stars position and velocity, convolved with the errors. Assuming a Gaussian
distribution, we have derived a value of the standard deviation for each stellar eccentricity. The mean
value of the standard deviation depends slightly on the eccentricity itself, and its equal to 0.14 at eccentricity$=$0.20
and 0.13 at eccentricity$=$0.80.

Furthermore, Figure~\ref{EccentricityMet} presents the stellar orbital eccentricities as a function of metallicity
for the thin and the thick disc sequences (lower and upper panels, respectively). On one hand, the thin disc sequence
presents a very similar eccentricity distribution through its whole metallicity range, with a majority of low eccentric orbits.
In addition, when only thin disc stars with a distance to the plane lower than 400~pc are considered (blue open circles in 
Fig.~\ref{EccentricityMet}), most of the thin disc stars with higher eccentric orbits disappear. In addition, this high eccentricity
tail of the thin disc suffers from a very low statistics and the relatively large errors in the derived eccentricities. 
On the other hand, the thick disc sequence seems to present highly eccentric orbits mainly in the metal-poor regime 
(\met$<-$0.6~dex). Additionally, the metal-rich part of the thick disc sequence shows very similar eccentricities 
to the thin disc one. A similar behaviour of the mean eccentricity distribution with metallicity has also been observed by
\citet{Adibekyan2013} in the solar neighbourhood. Finally, the derived slopes of the eccentricity dependence on
metallicity, $\Delta$e$/\Delta$\met, are -0.158$\pm$0.081~dex$^{-1}$ for the thick disc and -0.027$\pm$0.045~dex$^{-1}$ for the thin
disc. These values are also in agreement with those determined by \citet{Adibekyan2013} for the solar neighbourhood 
(-0.184$\pm$0.078~dex$^{-1}$ and -0.023$\pm$0.015~dex$^{-1}$ for the thick and the thin disc respectively).

\subsection{Maximum height above the plane}

\begin{figure}[th!]
\includegraphics[width=8.5cm,height=8cm]{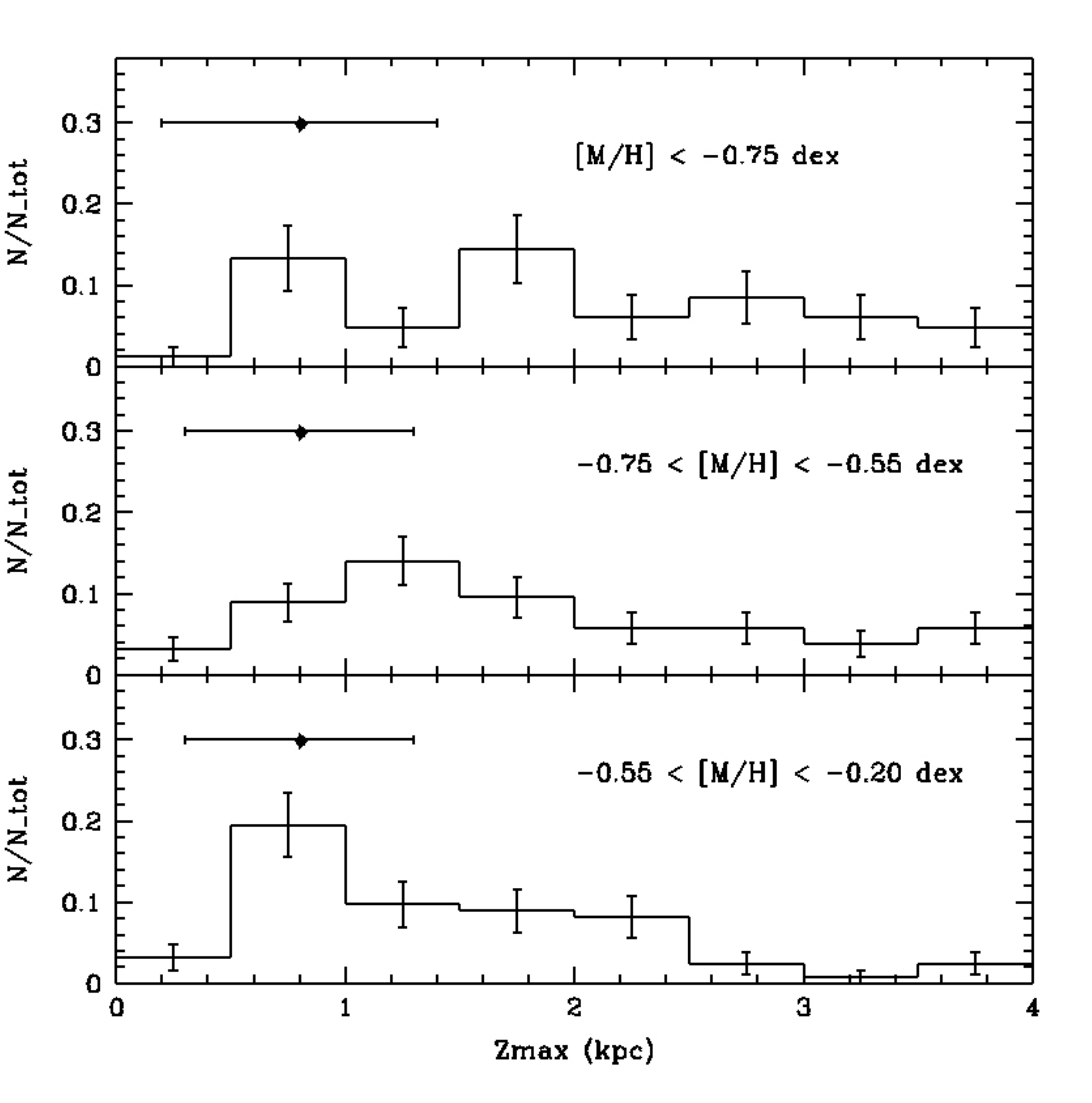}   
\caption{Normalized distribution of maximum heights above the Galactic plane (Zmax) for the stars in the thick disc sequence, divided in three metallicity intervals. The error bars for the Zmax are shown on the upper left corner of each panel. The Poisson uncertainties are also shown as the histogram's error bars.}
\label{Zmax}
\end{figure}

We have also studied the distribution of maximum heights above the Galactic plane, Zmax, for the stars in the
thick disc sequence. To this purpose, as shown in Fig.~\ref{Zmax}, we have plotted the normalized Zmax distribution for three
different metallicity intervals. The associated error to Zmax was calculated through the above described procedure, already
used for the eccentricity error, and it is of the order of 0.6~kpc.
  
As expected from the analysis of the mean distances to the Galactic plane in Sect.~5,
the Zmax distribution seems to peak at lower values as the metallicity increases. Since we are working with a colour/magnitude 
selected sample, the quantification of a gradient would need correction for the volume sampled as a function of metallicity.
Nevertheless, the observed Z and Zmax distributions seem to suggest a rather steep vertical metallicity gradient in the thick disk.
On the other hand, \citet{Katz11} found a low gradient of -0.068$\pm$0.009~dex~kpc$^{-1}$, derived without a chemical
selection of the thick disc in the \aabun \ vs. \met \ plane. On the other hand, \citet{Bilir12} derived a value of the vertical
metallicity gradient in the thick disc of -0.034$\pm$0.003~dex~kpc$^{-1}$, from a kinematically selected sample of RAVE stars. 
Finally, \citet{Ruchti11} found an iron abundance vertical gradient of -0.09$\pm$0.05~dex~kpc$^{-1}$ for metal poor thick disc stars.

\section{Radial gradients in rotation, metallicity and \aabun}

\begin{figure}[th!]
\includegraphics[width=8.5cm,height=8cm]{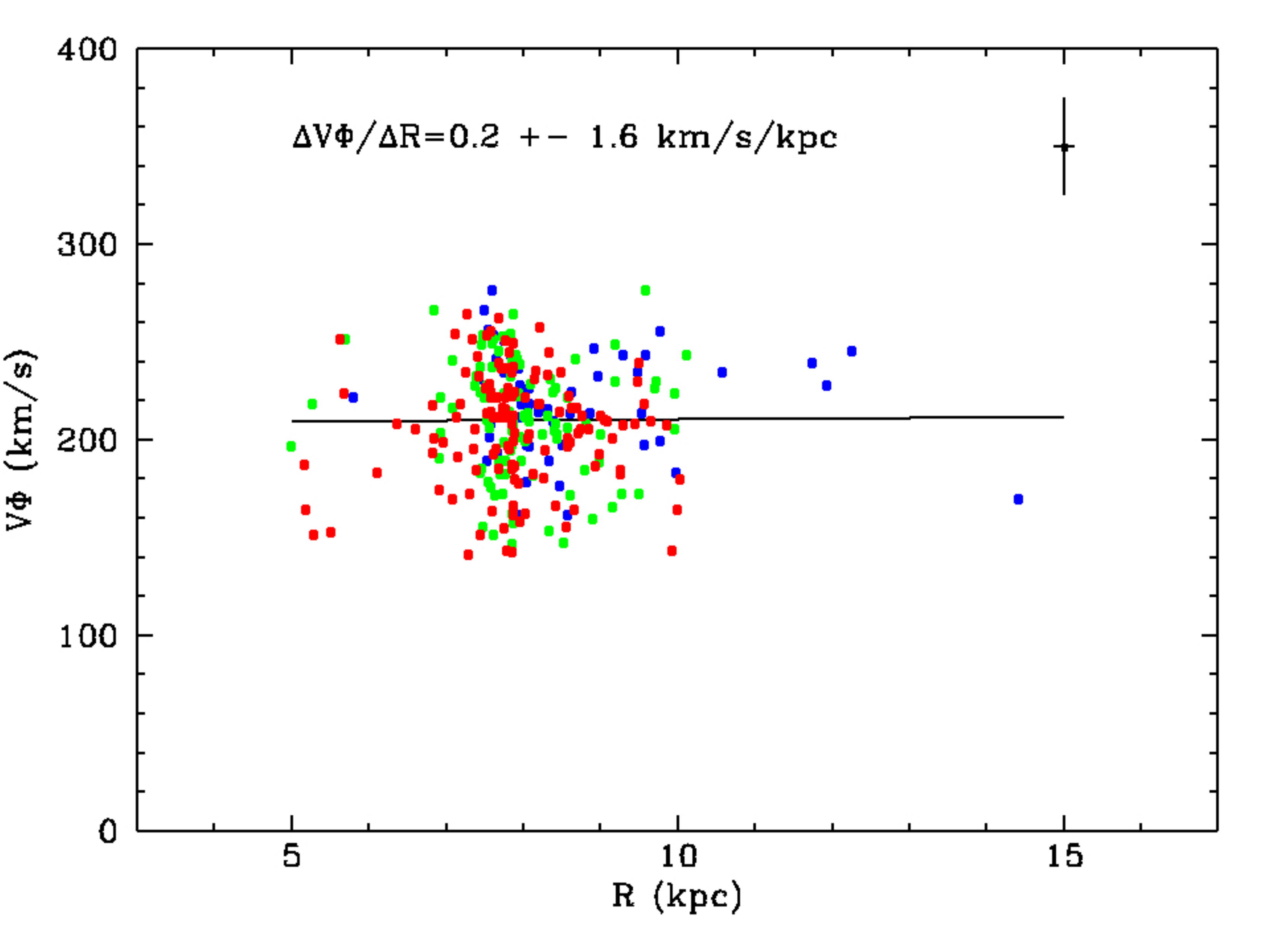}  
\caption{Rotational velocity gradients with Galactocentric radius for thin disc stars having $|$z$|<$700~kpc. 
The colour codes correspond to three 
metallicity intervals: \met$<$-0.5~dex (blue), -0.5$<$\met$<$-0.3~dex (green) and \met$>$-0.3~dex (red).}
\label{GalRot}
\end{figure}

\subsection{The thin disc rotational velocity behaviour with Galactocentric radius.}
First of all, as illustrated in Fig.~\ref{GalRot}, we examined a possible dependence of V$_{\Phi}$ with the
Galactocentric radius for the thin disc. To this purpose, we have selected from the thin disc stars of sub-sample {\it e} those 
within a distance to the Galactic plane lower than 700 pc.
We have also selected stars having errors in V$_{\Phi}$ smaller than 40~km/s for 7$<$R$<$9~kpc, up to 60~km/s
for 6$<$R$<$7~kpc and 9$<$R$<$10~kpc and 50$\%$ elsewhere. In addition,  we have performed a 2-$\sigma$ clipping
around the mean values of V$_{\Phi}$ in three metallicity intervals (\met$<$-0.5~dex, -0.5$<$\met$<$-0.3~dex and \met$>$-0.3~dex). 
The final sample contains 283 stars with a mean error in 
V$_{\Phi}$ of 23~km/s and a mean error in R of 130~pc. Due to the survey selection function for this first 
GES iDR1 release, most of the thin disc targets near the Galactic plane are within 5$<$R$<$11~kpc, with
very few objects in the outer parts of the disc. The resulting least squares fit of the data shows a flat behaviour of the
rotational velocity (with a slope of 0.2$\pm$1.6~km~s$^{-1}$~kpc$^{-1}$, where the uncertainty comes from the standard error
of the slope). 
For comparison, literature estimations of the Galactic rotation curve 
from individual velocities of 70 Carbon stars by \citet{Demers}, from gas clouds using the terminal 
velocities of the HI and CO lines \citep{Sofue} and from APOGEE data \citep{BovyApogee}, find an
approximately flat behaviour, as what is found for the rotational velocity with our data. 

In addition, we have tested the possible effect of the V$_{\Phi}$ correlation 
with metallicity (c.f. Sect.~6) on the derived  V$_{\Phi}$ vs. R relation. To this purpose, we have divided the above
described sample in three sub-samples for different metallicity intervals: \met$<$-0.5~dex, -0.5$<$\met$<$-0.3~dex and \met$>$-0.3~dex
(see Fig~\ref{GalRot}).
Then, three values of the slope were derived for each of the sub-samples. For all these metallicity intervals, the rotational velocity
stays flat within the error bar. Therefore, no measurable influence of the stellar rotation correlation with metallicity
seems to exist in the thin disc, within the studied galactocentric distance range. In agreement with Fig.~\ref{RotMetGrad}, 
the mean value of V$_{\Phi}$ decreases with metallicity, in the three considered metallicity intervals.
Finally, we have checked a possible trend of the V$_{\Phi}$ dispersion with R in the examined sample, 
finding a flat $\sigma_{\Phi}$ with
R in the range 7$<$R$<$10~kpc, for which the number of stars allows a robust enough measurement of the V$_{\Phi}$ dispersion.

\begin{figure}[th!]
\includegraphics[width=8.5cm,height=9cm]{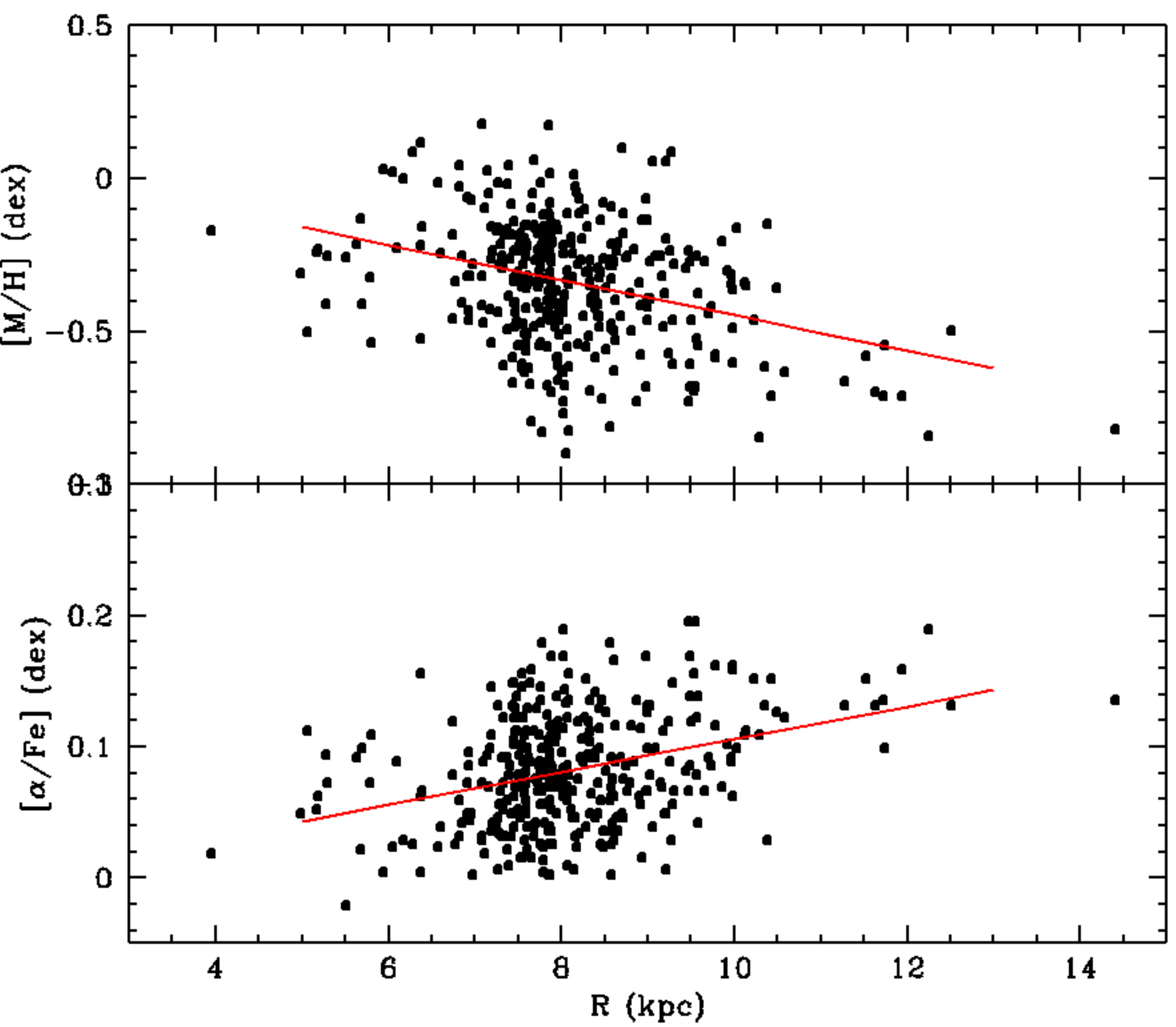}  
\caption{Global metallicity (upper panel) and \aabun \ (lower panel) as a function of the Galactocentric radius for thin disc stars
within 700~pc from the Galactic plane.}
\label{AbunGradThin}
\end{figure}

\subsection{Radial abundance gradients}

In order to derive the radial abundance gradients in the thin disc, we have used the sample of stars described in Section 7.1
and performed least square fits to the \met \ v.s. $R$ distribution and the \aabun \ v.s. $R$ distribution 
(c.f. Fig.~\ref{AbunGradThin}). We find a negative gradient of -0.058$\pm$0.008~dex~kpc$^{-1}$ for \met \ and a very small positive 
gradient of 0.012$\pm$0.002~dex~kpc$^{-1}$ for the \aabun. As in the previous section, the uncertainties come from the standard
error of the slope, derived by the least square linear regression. 

The value of the metallicity gradient is a matter of debate, ranging 
from -0.06 to -0.1~dex~kpc$^{-1}$. Our measured value is therefore in agreement within the literature intervals,
as the most recent studies using Cepheid stars \citep[around -0.06~dex~kpc$^{-1}$, see][and references therein\footnote{Measured with iron
abundances}]{Lemasle13} and dwarfs stars in the Corot
mission fields \citep[-0.053$\pm$0.011~dex~kpc$^{-1}$][]{Gazzano13}. Measurements based on HII regions\footnote{Measured with oxygen, nitrogen, and sulphur abundances.} range from 
-0.046$\pm$0.009~dex~kpc$^{-1}$ to -0.071$\pm$0.010~dex~kpc$^{-1}$ in the optical and -0.041$\pm$0.009~dex~kpc$^{-1}$ to 
-0.085$\pm$0.010~dex~kpc$^{-1}$ in the infrared \citep{Rudolph}. Our metallicity gradient estimation is also in agreement with the
\citet{Cheng12} analysis of SEGUE data, who find -0.14~dex~kpc$^{-1}$ for 0.15~kpc$<$Z$<$0.25, -0.082~dex~kpc$^{-1}$ for 0.25~kpc$<$Z$<$0.50, and -0.024~dex~kpc$^{-1}$ for 0.50~kpc$<$Z$<$1.0, selecting low$-\alpha$ stars in their sample.
Finally, observations of open clusters find steeper gradients for the inner regions \citep[$\sim -0.07$~dex~kpc$^{-1}$ for R$<$12~kpc, c.f.][]{Andreuzzi, Magrini}, than for the whole covered radial range \citep[$\sim -0.05$~dex~kpc$^{-1}$,][]{Andreuzzi, Ricardo11}. 
In addition, it is worth noticing that the different above mentioned indicators trace the gradient at different times, and the issue of 
the gradient evolution with time is indeed still open.

On the other hand, the very small positive gradient in the \aabun \ abundance ratio 
seems not to be influenced by the cut in vertical distance to the Galactic plane, as its value does
not change if the cut is done at $|z|<$400~pc instead of 700~pc. This could be the case if a contamination by thick disc stars was
affecting the radial gradient measurement. However, the presence of numerous metal-poor thin disc stars
with \met$<-$0.5~dex in our sample, influences the measurement, steepening the gradient, as those stars present high values of \aabun \ and
dominate the outer regions. If real, this slightly positive gradient in \aabun \  could be in agreement with the measurements of 
\citet{Yong05} and \citet{Ricardo11} for open clusters, \citet{Carney} for field giants, and \citet{Yong06} for Cepheids, who report
an enhancement of \aabun \ in the outer disc. In particular, the result for the Cepheids of \citet{Yong06} comes from stars in
the metal-poor regime (-1.0$<$\met$<$-0.3~dex). Nevertheless, the individual $\alpha$-element abundances of Cepheids reported 
by \citet{LuckLambert11} in the range $\sim$4-15~kpc are compatible with a rather small positive gradient in $[$Mg/Fe$]$, whereas
\citet{Lemasle13} report a $[$Si$/$Fe$]$ gradient close to zero and a very small gradient in $[$Ca$/$Fe$]$ of 0.016~dex/kpc (assuming 
the well accepted metallicity gradient for Cepheids of -0.06~dex/kpc), close to their measurement errors.

Moreover, we have checked the influence of the data spatial distribution in the derived gradients. 
For that purpose, it is necessary to test the robustness of the two derived gradients on the
inhomogeneity of the survey in the R and Z distribution. As already explained in Section~5, 
our total sample is dominated by stars further away from the galactic plane in the inner parts of the disc with
respect to the typical heights probed in regions outside of the solar radius. However, thanks to the restriction
of the thin disc sample to Z values smaller than 700~pc, the above mentioned correlation between Z and R practically
dissapears: the mean distance to the Galactic plane is $\sim$450~pc at R$\sim$6~kpc and $\sim$300~pc at R$\sim$12~kpc.
This small trend can not be responsible for the measured gradients in metallicity and \aabun, especially, after the
chemical selection of the stars from the thin disc sequence of the \met \ vs. \aabun \ plane. In addition, in the case
of the \aabun \ gradient, the small negative trend of Z with R excludes a higher spurious thick disc contamination 
in the outer regions with respect to the inner ones, as we would expect if the \aabun \ gradient was artificially
created by a bad thin-thick disc separation in the outer regions.
Therefore, as a consequence of the two gradients in \met \ and \aabun, the outer parts of the thin disc sounded by the
present sample of GES data (up to $R\sim$12~kpc) appear more metal-poor and, possibly, slightly richer in 
\aabun \ than the inner regions.

\begin{figure}[th!]
\includegraphics[width=8.5cm,height=9cm]{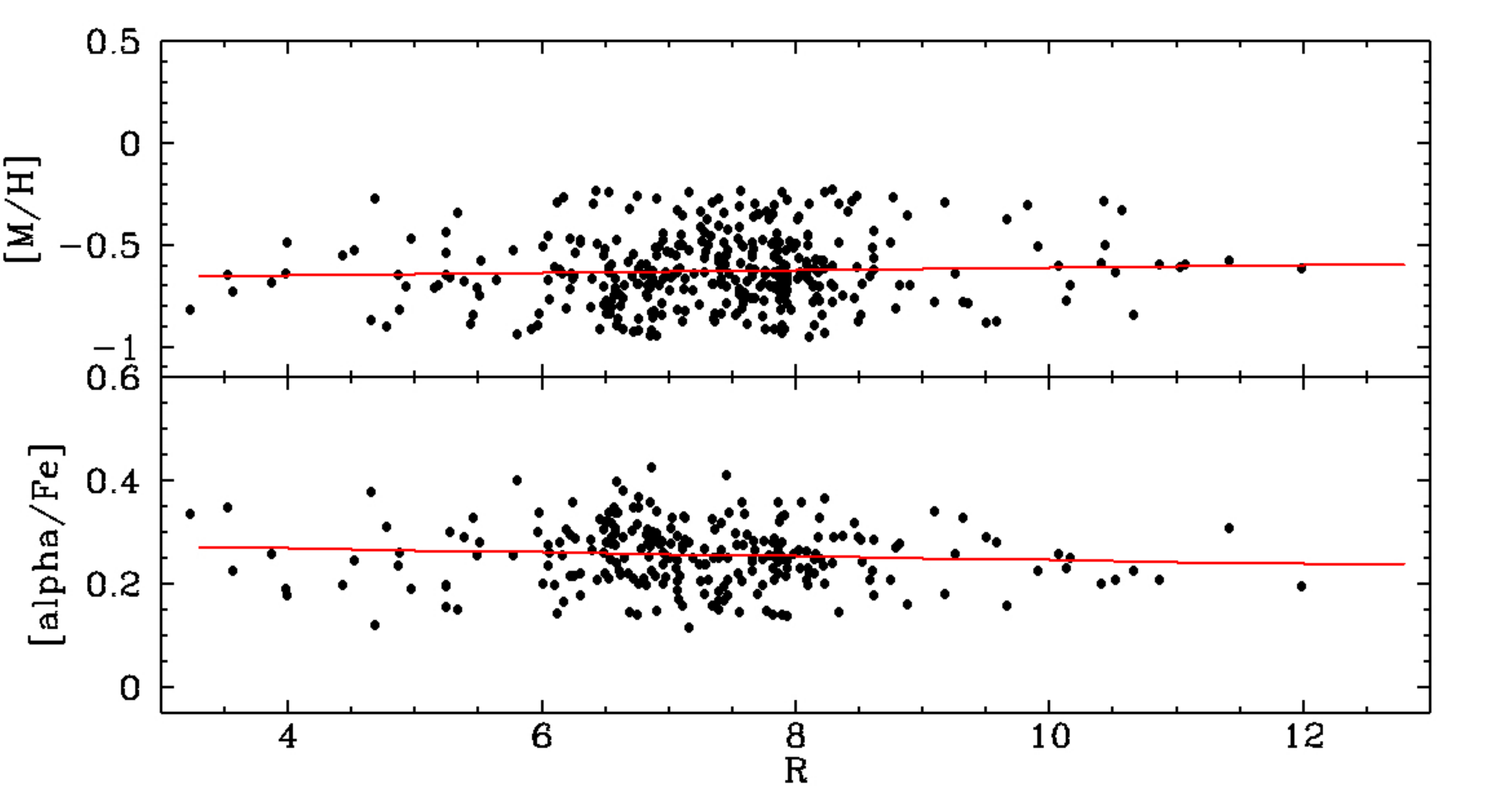}  
\caption{Global metallicity (upper panel) and \aabun \ (lower panel) as a function of the Galactocentric radius for thick disc stars
with Z$>$700~pc from the Galactic plane.}
\label{AbunGradThick}
\end{figure}

Finally, we have also analysed the radial abundance gradients in the thick disc population (c.f. Fig~\ref{AbunGradThick}), 
selecting stars in the thick disc sequence of the \aabun \ vs. \met \ plane, with Z distances 
higher than 700~pc. The cut in Z limits the contamination by thin disc stars and possible
problems with the inhomogeneus sampling in R and Z. The derived slopes, both for \met \ and \aabun \ 
for thick disc stars, are within the corresponding error bars (0.006$\pm$0.008~dex and -0.004$\pm$0.003~dex respectively). 
Therefore, our data suggest a flat distribution with galactocentric radius, both for \met \ and \aabun \ for thick disc stars.
Our result for the \met \ gradient is in agreement with the \citet{Cheng12} flat gradient for their selected
high-$\alpha$ stars from SEGUE data.

\section{Summary and discussion}
We have used the atmospheric parameters, \aabun \ abundance ratios  and radial velocities, determined from 
the Gaia-ESO Survey GIRAFFE spectra of FGK-type stars (first nine months of observations), to present
the kinematical and chemical characterisation of the observed stellar populations. First, we have
chosen to identify the thin to thick disc separation in the \aabun \ v.s. \met \ plane, thanks to
the presence of a low-density region in the number density distribution. This separation
is in agreement with the recently proposed one by \citet{Adibekyan2013} from a sample of solar neighbourhood
stars, and the ones observed before by \citet{Fuhrmann04}, \citet{Reddy06} and \citet{Bensby07}, among others.

Furthermore, we have studied the distribution of the distances to the Galactic plane, in the \aabun \ v.s. \met \ plane,
with particular attention to the chemically defined thin and thick disc populations. Regarding the thick disc, our results
show that the stars lay in progressively thinner layers along the sequence (as metallicity increases and \aabun \ decreases).
This finding could be in agreement with  the \citet{Haywood2013} proposal of a thick disc forming stars in thinner 
and thinner layers for 4-5 Gyrs, also predicted by the dissipational settling formation scenario of \citet{Burkert92}. 
Similarly, \citet{Bovy12Mono}  already pointed out the scenario of an evolution in gradually thinner layers (as suggested by the
dependence of their scale height distribution on \aabun), although no clear separation between the thin
and the thick disc sequences appears in their data. In addition, the recent hydrodynamic simulations of \citet{Bird13}
and \citet{Stinson13} also show discs in which the younger stars have progressively shorter vertical scale heights.
On the other hand, the thin disc sequence presents a constant value of the mean distance
to the Galactic plane at all metallicities. As a consequence, the thick to thin disc transition in Z values appears
more abrupt in the low metallicity regime. 

Concerning the kinematics, the mean rotational velocities found for the thick and the thin disc sequences are in agreement with
the canonical ones (V$_{\Phi}=$176$\pm$16~km/s and V$_{\Phi}=$208$\pm$6~km/s). In addition, the mean Galactic rotation
also seems to increase progressively along the thick disc sequence, as the metallicity increases. Our data also confirm the already observed
correlations between V$_{\Phi}$ and \met \ for the two discs. In the case of the thick disc sequence, the mean value of
the $\Delta$V$_{\Phi}$/$\Delta$\met \ is 43$\pm$13~km~s$^{-1}$~dex$^{-1}$, is in agreement with the previous findings
of \citet{Lee2011b}, \citet{Kordopatis11b} and \citet{Spagna2010}. This would confirm the disagreement with the SDSS result
of \citet{Ivezic}, based on photometric metallicities. Moreover, we have explored the possible influence of contaminating
stars from the thin disc in the analysis of the correlation. If only the stars with the higher \aabun \ values at each
metallicity bin are considered, the derived $\Delta$V$_{\Phi}$/$\Delta$\met \ is 64$\pm$9~km~s$^{-1}$~dex$^{-1}$. As a consequence, 
this steeper value of the gradient, cleaned from the influence of possible thin disc contaminants could be more characteristic 
of the real correlation for the thick disc. For the thin disc, a negative gradient of 
$\Delta$V$_{\Phi}$/$\Delta$\met$=$-17$\pm$6~km~$s^{-1}$~dex$^{-1}$ is found. Consequently, the difference in the characteristic
rotational velocities between the thin and the thick disc are again accentuated in the low metallicity regime, as for the 
mean distances to the Galactic plane. Furthermore, the analysis of the velocity dispersions for the chemically separated 
thin and thick disc stars confirms a higher dispersion for the thick disc population than for the thin disc one, as expected, 
for both V$_{\Phi}$ and V$_{z}$. Further, the stars in the thick and the thin disc sequences present also different distributions
of their stellar orbit eccentricities, with the thick disc stars presenting a broader distribution peaked at higher eccentricities
than the thin disc one.

Finally, we have analysed the properties of the thin and thick discs as a function of the Galactocentric radius. 
For the thin disc, we estimate a flat behaviour of the thin disc rotational velocity,
a metallicity gradient equal to -0.058$\pm$0.008~dex~kpc$^{-1}$ and, a very small positive \aabun \ gradient.
For the thick disc, no gradients in \met \ and \aabun \ are found.

The general picture emerging from our analysis is that of a thick disc that lays in progressively thinner and thinner layers
as the metallicity increases and the \aabun \ decreases with time (until about \met~$=-0.25~$dex and \aabun~$=0.1~$dex). 
During this settling process, the thick disc rotation
increases progressively and, possibly, the azimuthal velocity dispersion decreases. No measurable metallicity or
\aabun \ gradients with galactocentric distance seem to remain for the thick disc. At about \met~$=-0.25~$dex, the
mean characteristics of the thick disc in \aabun, vertical distance to the plane, rotation and 
rotational dispersion are in agreement with (or just slightly different than) that of the thin disc stars of the same metallicity,
suggesting a possible connection between the two populations at a certain moment of the disc evolution.
This is in agreement with the \citet{Haywood2013} scenario, in which the inner thin disk (R$<$10kpc) inherited from the chemical
conditions left at the end of the thick disk phase (the observed metal-rich end of the thick disc sequence in the
\aabun \ vs. \met \ plane), some 8 Gyr ago. In their view, there may be an age gap in the star formation possibly 
causing the gap in \aabun \ enrichment, after which, star formation proceeds in a thin disk.

For the thin disc, no clear gradient of the mean vertical distance to the plane with the stellar metallicity is observed. 
On the contrary, the stars in the metal-poor thin disc sequence tail seem to have a mean Galactic rotation velocity higher than those
in the metal-rich part. Moreover, if confirmed, the existence of a small positive gradient in \aabun \ with
R leads to a thin disc that would be slightly more metal-poor and \aabun-rich in the outer parts. 
These two characteristics, similar to those of the metal-poor end of the thin disc sequence in the 
\aabun \ v.s. \met \ plane, would favour the picture, suggested by \citet{Haywood2013} and \citet{Haywood2006},
of this metal-poor thin disc being formed in the outer parts.

Contrary to the similarities between the thin and the thick discs around \met$\sim-$0.3dex, the stars in the metal-poor
overlapping regime are more clearly separated in their mean physical properties (chemistry, kinematics, possibly orbital
eccentricities). This separation seems to be confirmed by the gap (or low-density region) in the \aabun \ vs. \met.

On the other hand, the tight bimodal distribution of \aabun \ at every value of \met \ poses severe constraints to models predictions.
Recent attempts to define an age-metallicity relation for stars currently in the solar neighbourhood \citep[e.g.][]{Haywood2013, Mishenina13}
hint at an evolutionary path in which a very wide range of ages corresponds to a single \met \ value.
\citet{Haywood2013} results (c.f. their Fig.~8) show that at a value of about \met$=$-0.5~dex stars have an age range of around 10~Gyr. 
On the other hand, the present work and other studies focussed in the solar neighbourhood \citep[e.g.][]{Adibekyan2013} 
show that the distribution of \aabun \ at \met$=$-0.5~dex appears narrowly bimodal, with two essentially unresolved but quite discrete 
values of \aabun \ being found. As a consequence, although stars with \met$\approx$-0.5~dex formed over 10Gyr, they seem to be formed 
with only two values of \aabun \. This is a quite remarkably severe constraint that can be analysed in the light of Galactic models in which radial migration \citep[e.g.][]{Sellwood02, Schonrich2009, Schonrich2009b, Roskar08}, and churning, is a major factor.

 \citet{Schonrich2009b} claim that a bimodal, although continuous, distribution of \aabun \ can naturally come out as a consequence 
of the standard assumptions about star formation rates and metal enrichment, with no need to invoke breaks in the Galaxy's star formation 
history  or accretion events. However, the tightness of the well populated observed thick and thin disc sequences in the \aabun \ vs. \met \ 
plane, with possibly only two values of \aabun \ for a given metallicity, poses severe constraints.

If radial migration occurs, one can suppose an evolutionary path in which enrichment in the high-$\alpha$ branch  happened rapidly, up to metallicities close to \met$=-$0.25~dex. This should occur across a large enough
radial range of the proto-galactic disk. In fact, to preserve the tightness of the high-$\alpha$ sequence, the stars should had formed
in a narrow time interval from radially well-mixed interstellar medium (ISM) gas, through all the
galactocentric radii which generated stars that are today at a given R (ex. in the solar neighbourhod). Then, to explain 
the low-$\alpha$ sequence, one can postulate a significant gas accretion, reducing the disc ISM metallicity to, for instance, 
-0.6~dex. This should be followed by disc evolution in which large scale gas flows 
\citep[as in fountains, c.f.][and references therein]{Fraternali13} retain a tight \aabun-\met \ 
correlation, over the radial region which today populates the solar neighbourhood. In addition, this scenario should, somehow, 
also conciliate the higher rotation values observed in the metal-poor thin disc regime. 
Othewise, other possibility to explain the observed \aabun \ vs. \met \ distributions, is to suppose 
that strong radial migration in the sense of churning did not occure \citep[see also the discussion in][]{Haywood2013}.


Finally, this study also shows the importance of precise chemical abundance measurements to disentangle the stellar population 
puzzle of the Galactic disc. The stellar chemical patterns can guide the definition of useful stellar sub-samples to unveil the 
evolutionary paths in the
disc formation history and to give clearer constraints to the models. Moreover, in the near future, the Gaia mission of the European Space
Agency will allow precise distances and ages estimations for all the analysed stars. The results presented here, based only on the first months of the GES 
observations, confirm how crucial are today large high-resolution surveys outside the solar neighbourhood for our understanding
of the Milky Way history.



\begin{acknowledgements}
 Based on data products from observations made with ESO Telescopes at the La Silla Paranal Observatory under programme ID 188.B-3002.
The results presented here benefit from discussions held during the Gaia-ESO workshops and conferences supported by the ESF (European Science Foundation) through the GREAT Research Network Programme. A. Recio-Blanco, P. de Laverny and V. Hill acknowledge the the support of the French Agence Nationale de la Recherche, under contract ANR-2010-BLAN- 0508-01OTP, and the "Programme National de Cosmologie et 
Galaxies" (PNCG) of CNRS/INSU, France. 
This work was partly supported by the European Union FP7 programme through ERC grant number 320360 and by the Leverhulme Trust through grant RPG-2012-541. We acknowledge the support from INAF and Ministero dell'Istruzione, dell' Universita' e della Ricerca (MIUR) in the form of the grant "Premiale VLT 2012". A. Helmi acknowledges financial support from ERC-Starting Grant, nr. GALACTICA-240271. Finally, we are sincerely grateful to the extremely careful anonymous referee, who undoubtedly enhanced the value of this paper.

\end{acknowledgements}

\bibliographystyle{aa.bst}
\bibliography{biblio}

\end{document}